# A comparative study of auxiliary intercalating agents on thermal conductivity of expanded graphite/polyetherimide composite

Fatema Tarannum, Swapneel Danayat, Avinash Nayal, Rajmohan Muthaiah, Roshan Sameer Annam, Jivtesh Garg

School of Aerospace and Mechanical Engineering, University of Oklahoma, Norman,73019, USA

**Abstract**

In this work, we have comprehensively studied the effect of auxiliary intercalating agents on the thermal conductivity of expanded graphite (EG) polymer composites. We report an ultra-high enhancement of 4030% in thermal conductivity of polyetherimide/graphene nanocomposite ($k$ = 9.5 Wm$^{-1}$K$^{-1}$) prepared through the use of EG with hydrogen peroxide ($H_2O_2$) as an auxiliary intercalating agent at 10 wt% compositions ($k$ of pure polyetherimide ~ 0.23 Wm$^{-1}$K$^{-1}$). This ultra-high thermal conductivity value is found to be due to an EG-mediated interconnected graphene network throughout the composite, establishing a percolative environment that enables highly efficient thermal transport in the composite. Comparative studies were also performed using sodium chlorate ($NaClO_3$) as an auxiliary intercalating agent. At 10 wt% composition, $NaClO_3$ intercalated EG was found to lead to a smaller enhancement of 2190% in $k$ of composite. Detailed characterization performed to elucidate this advantage, revealed that hydrogen peroxide led to primarily edge oxidation of graphene sheets within EG, leaving the basal plane intact, thus preserving the ultra-high in-plane thermal conductivity of ~ 2000 Wm$^{-1}$K$^{-1}$. Sodium chlorate, on the other hand, led to a higher degree of oxidation, with a large number of oxygen groups on the basal plane of graphene, dramatically lowering its in-plane thermal conductivity. Thermal diffusivity of $H_2O_2$ prepared EG paper was measured to be 9.5 mm$^2$/s while that of $NaClO_3$ case was measured to be 6.7 mm$^2$/s, thus directly confirming the beneficial impact of $H_2O_2$ on $k$ of graphene itself. This study is the first to address the role of intercalating agents on $k$ of expanded graphite/polymer composites and has led to the discovery of $H_2O_2$ as an effective intercalating agent for achieving ultra-high thermal conductivity values.

## 1. Introduction

Graphite comprises of multiple layers of graphene which are stacked along the c-axis by weak van der Waals force with an interlayer spacing of 3.35 Å[1]. Due to the weak van der Waals force and the interlayer spacing, the insertion of atoms, ions, and molecules between the graphene layers can quickly initiate the intercalation process. Graphite intercalated compound (GIC) is the form of graphite with intercalated chemical species such as atoms, ions, and molecules between the graphene layers[2,3] as shown in Figure 1a. GIC has been studied for various applications including- superconductors, fuel cells, battery cells, heterogeneous catalysts, electrodes, hydrogen storage, and polarizers[4-7]. GICs are also of interest for obtaining large lateral size[8-10] and single/bilayer/few layers of graphene using liquid-phase exfoliation[1,11].

Concentrated $H_2SO_4$ serves as the acidic environment and has been the most common intercalating agent for the preparation technique of GIC; however, the intercalation process also requires an anodic or strong oxidizing agent. According to Cai *et al.*[12], even though the van der Waals force between the graphene layers is weak, it is still strong enough to make the intercalation process difficult. Oxidizing agents coupled with $H_2SO_4$ reduce these forces significantly, enabling the interlayer separation to be increased, which leads to a high degree of intercalation of GIC[12,13]. Intercalation degree is directly related to a parameter known as staging index which is equal to the number of graphite layers between two intercalated regions[14]. Several studies emphasize the role of staging index of GIC[15,16] in understanding the intercalation process.

According to the Rudorff model, a single layer, bilayer, or tri-layer graphene can be alternated regularly with intercalated species in stages I, II, or III. In stage-I of the intercalation process, each graphene sheet is separated from the others by intercalated compounds, and in stage II, GIC has two adjacent graphene sheets contained between intercalated compounds[17] as shown in Figure 1b.

GIC possesses a well-stacked graphitic structure containing acceptor, donor, or neutral type intercalant species. An acceptor type GIC is produced when electronegative species accept an electron and form an ionic bond with the π-electron network[16] denoted by $C_x$.

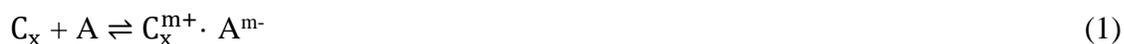
$$C_x + A \rightleftharpoons C_x^{m+} \cdot A^{m-} \tag{1}$$

Here A accepts m electron from the π-electron network. A donor type GIC is produced when an electron is donated to the network[16,18]; this happens with metal atoms as shown in the following equation.

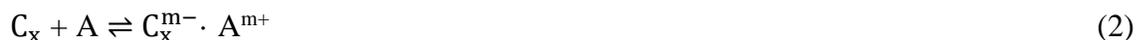
$$C_x + A \rightleftharpoons C_x^{m-} \cdot A^{m+} \tag{2}$$

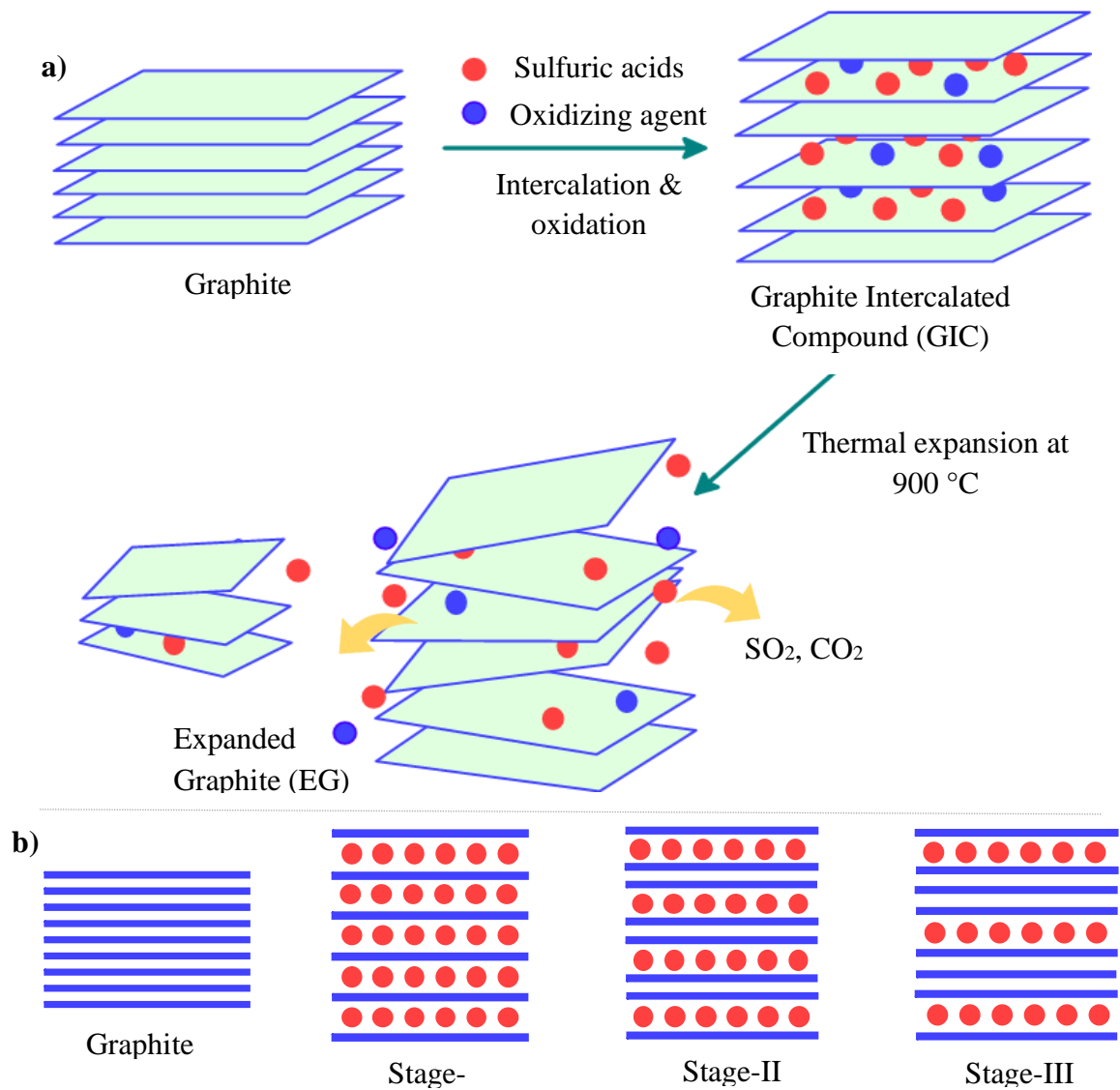

**Figure 5.1** a) Preparation of graphite intercalation compound (GIC) in the presence of intercalants (acids & oxidizing agents), b) schematic illustration of different stages of intercalation.

The intercalation process allows simultaneous oxidation or reduction of starting graphite sheets. Through this method, the positive or negative charges of the acceptor or donor type GIC cause repulsion between the graphite layers that increase their relative distance, and the intercalating agents can enter in between the graphite gaps. Donor and acceptor type intercalants cause the carbon-carbon bonds in the bonding layers to have stage-dependent stiffening nature for acceptors or softening for donors, respectively, due to the electron transfer process[19]. Another main structural characteristic of GIC is strong intraplanar binding and weak interplanar bonding, which causes the graphite and intercalant layers to stay separated and maintain the inherent properties of graphite.

A class of graphite intercalated compounds known as graphite bisulfates, are an acceptor type intercalated compound and can be synthesized using sulfuric acid and oxidizing agents such as concentrated nitric acid, potassium permanganate, potassium dichromate and chromium trioxide[20,21].

Synthesis of graphite bisulfate is obtained using the following reaction scheme[22]:

$$24nC + O_x^z + mH_2SO_4 \rightarrow C_{24n}^+ \cdot HSO_4^- \times (m-1) H_2SO_4 + HO_x^{(z-1)} \qquad (3)$$

where $O_x$ is the oxidizing agent, C is the carbon atoms in the graphite, n, and m are the numbers of moles of C and $H_2SO_4$ respectively. Intercalation of graphite produces intercalated graphite bisulfate compound. Simultaneously, oxidizing agents cause the anionic and molecular insertion between the carbon layered structure along with the formation of $C_n^+$ through their high redox potential. Upon synthesis, GIC contains carbon, oxygen, and sulfur with different ratios. Due to the insertion of intercalated species and oxygen groups, GIC has huge potential to expand significantly by thermal heating, along the graphite c-crystallographic direction because of weak adhesive forces between adjacent graphene layers. Thus, GIC provides an efficient route to obtain expanded graphite (EG). Camino *et al.* have proposed that the expansion occurs due to the formation of carbon dioxide ($CO_2$), sulfur dioxide ($SO_2$), and water vapor ($H_2O$) produced by the reaction (5), yielding expanded graphite[23,24] (Figure 1a).

$$C + 2H_2SO_4 \rightarrow CO_2 + SO_2 + H_2O \qquad (4)$$

The properties of expanded graphite have been shown to be highly dependent on the GIC production processes related to intercalation, chemical oxidation, and electrochemical intercalation[25-27]. The characteristics of EG are substantially influenced by the sulfuric acid concentration, oxidizing agent, and intercalation period used in the intercalation process.

Numerous GICs with different intercalants have been synthesized and reported. In recent years, after discovering the fascinating properties of graphene, and the higher-temperature superconductivity in $CaC_6$[28], GICs started gaining renewed interest[29]. Significant research has been performed on the synthesis process and characterization of graphite intercalated compound. The intercalation process is complicated, and studies are still ongoing to better understand it. Hong *et al.* developed a simple method for producing graphene from sulfuric acid intercalated graphite oxide by fast reduction and expansion exfoliation at temperatures just above 100 °C in ambient air[30]. Salvatore et al.[31] explored the morphology and intercalation effect of intercalated graphite bisulfate synthesized using various oxidizing agents and sulfuric acid. Parvez et al.[32] investigated electrochemical exfoliation using several organic salts (($NH_4$)$_2$$SO_4$, $Na_2SO_4$, $K_2SO_4$) to create GICs using sulfate ions, and found that the EG, produced has better electrical properties including 11 Ω sq$^{-1}$ conductivity of graphene films and high yield exfoliation[32].

EG filler has recently attracted a lot of interest in polymer composite applications. While integrated with the polymer, EG filler improves thermal characteristics significantly. Huang et al.[33] found that two-step intercalation with hydrogen peroxide and phosphoric acid improved the flame retardancy of expandable graphite (EG). They found that 30 wt% EG composition EG/ethylene-vinylacetate copolymer (EVM) matrix composite had a high limited oxygen index (LOI) of 30.4% [33]. Hou et al. used hydrogen peroxide and sulfuric acid to make an exfoliated or expanded graphite film with an in-plane thermal conductivity of 575 Wm$^{-1}$K$^{-1}$ at a density of 2 gcm$^{-3}$. Liu et al.[34] reported thermal conductivity of 12.95 Wm$^{-1}$K$^{-1}$ of 3D-EG/ polydimethylsiloxane (PDMS) composite containing 31.9 wt% EG and graphene oxide (GO). Wang et al. demonstrated superior electrical properties of 1719 Sm$^{-1}$ of chemically expanded graphite poly (methyl methacrylate) (PMMA) composite through interlayer polymerization at 10 wt% EG filler content. Kuan et al. compared different preparation techniques to observe the flame retardant property and found that hydrothermally $H_2O_2$ processed EG high-density polyethylene composite showed superiority over other procedures[36]. However, no work has addressed the role of intercalating agents in the thermal conductivity enhancement of expanded graphite polymer composites.

To synthesize intercalated graphite, two distinct oxidizing agents, namely, hydrogen peroxide ($H_2O_2$) and sodium chlorate ($NaClO_3$) are used in this work, and the influence of thermally expanded graphite (EG) on the thermal characteristics of EG polymer composites is investigated in this work. Theoretical studies have revealed that Na ions possess strong intercalating power between graphene layers, resulting in lower stage number at a higher degree of oxidation[37]. Recently Kang et al.[38] theoretically studied the role of alkali metal sodium (Na) on the intercalation behavior of expanded graphite oxide with different amounts and ratios of oxygen functional groups. Wang et al. [39] also reported that oxidation leads to enlarged interlayer spacing of graphite to accommodate sodium ions and microchannels of oxidized graphite significantly help in Na ion diffusion. Intercalation behavior and effect on the thermal properties of EG using such alkali metal base and strong oxidizer, namely, $NaClO_3$ will be investigated here. On the other hand, Vittore et



*al.*[40] developed a simple approach for preparing edge oxidized graphite utilizing $H_2O_2$ treatment at 60 °C and also claimed an additional benefit of eliminating amorphous carbon fraction from starting graphite. Edge oxidized graphite can lead to superior thermal and electrical properties to the polymer composites. In this work, we utilize another intercalation route of $H_2O_2$ to prepare expanded graphite from intercalated graphite bisulfate. The role of auxiliary intercalating agents in modifying the thermal properties of expanded graphite is explored in this work with the goal of developing high thermal conductivity polymer/expanded-graphite nanocomposites.

Polyetherimide (PEI) has significantly low thermal conductivity of 0.23 $Wm^{-1}K^{-1}$. Thermally expanded graphite (EG) produced from GICs functions as a useful 3D carbon filler for the thermosetting polymer, PEI. The thermal conductivity of an EG/polymer composite made of $H_2O_2$ expanded graphite (EG-$H_2O_2$) was found to be significantly higher than that of $NaClO_3$ expanded graphite (EG-$NaClO_3$). Compared to pure PEI, the *k* value of EG-$H_2O_2$/PEI composite for $H_2O_2$ intercalation increased by ~ 4030%, whereas EG-$NaClO_3$ fillers increased the *k* value by ~2190% at 10 wt% expanded graphite filler concentration. We have also made compressed EG sheets out of EG-$H_2O_2$ and EG-$NaClO_3$ to compare the thermal conductivity of graphene paper attained via the two intercalation processes addressed above. In this study, the impact of oxidizing agent concentrations, reaction duration, and interlayer spacing on EG structure, chemical composition, and structural integrity of GIC and EG is thoroughly examined. To perform the investigation and provide evidence to demonstrate the remarkable improvement of EG-$H_2O_2$/PEI composites, characterization methods such as Raman, XPS, XRD, and FE-ESEM were utilized. The solvent casting approach was used to prepare the composite as it helps to preserve the EG structure during the composite preparation.

## 2. Experimental Sections

### 2.1 Materials

Natural flake graphite (-10 mesh graphite, 99.9%) and N, N-dimethylacetamide (DMAC) were purchased from Alfa aesar[41], US. Sulfuric acid ($H_2SO_4$, 95–98%), hydrogen peroxide ($H_2O_2$, 30%), sodium chlorate ($NaClO_3$, 99%), and polyetherimide (PEI pellets, melt index 18 g/10 min) were purchased from Sigma Aldrich[42].



## 2.2 Synthesis of GIC-$H_2O_2$ & EG-$H_2O_2$ using Intercalation Route I

10 mesh graphite was intercalated using $H_2SO_4$ and $H_2O_2$ under mechanical stirring in the intercalation route I. Initially, a cold-water bath was used for 10-15 min to run the reaction then we ran the reaction at room temperature. 20 ml of $H_2SO_4$ was added to the flask and slowly cooled down below 20 °C using a cold-water bath. Then 2 g graphite was added to the $H_2SO_4$ solution, followed by a very slow addition of $H_2O_2$. 300 ml cold deionized water was added to the mixture slowly at the end of the reaction time. Then we filtered and separated the particles from the acidic solution. We dried the filtered graphite at 60 °C for 24 h to obtain the intercalation compound GIC, denoted as GIC-$H_2O_2$. Then, the EG particles, named EG-$H_2O_2$, were obtained from GIC-$H_2O_2$ through rapid heat treatment at 900 °C. To achieve the highest thermal properties, we optimized the amount of $H_2SO_4$ and $H_2O_2$ using different quantities of $H_2SO_4$ and $H_2O_2$ in the reaction. Performed reaction conditions and obtained sample names are given in Table 1.

**Table 1** Chemicals and their compositions used for the synthesis of GIC-$H_2O_2$ and EG-$H_2O_2$

| Sample name (GIC & EG) | 10 Mesh Graphite (g) | Amounts of reactants used | | Volume ratio ($H_2SO_4$: $H_2O_2$) | Time |
|---|---|---|---|---|---|
| | | $H_2SO_4$ (mL) | $H_2O_2$ (mL) | | |
| GIC-$H_2O_2$ 1 & EG-$H_2O_2$ 1 | 2 | 20 | 6 | 3.33:1 | 30 min |
| GIC-$H_2O_2$ 2 & EG-$H_2O_2$ 2 | 2 | 40 | 6 | 6.67:1 | 30 min |
| GIC-$H_2O_2$ 3 & EG-$H_2O_2$ 3 | 2 | 60 | 6 | 10:1 | 30 min |
| GIC-$H_2O_2$ 4 & EG-$H_2O_2$ 4 | 2 | 20 | 2 | 10:1 | 1 h |
| GIC-$H_2O_2$ 5 & EG-$H_2O_2$ 5 | 2 | 20 | 4 | 5:1 | 1 h |
| GIC-$H_2O_2$ 6 & EG-$H_2O_2$ 6 | 2 | 20 | 6 | 3:33:1 | 1 h |



## 2.3 Synthesis of GIC-NaClO$_3$ & EG-NaClO$_3$ using Intercalation Route II

In intercalation route II, H$_2$SO$_4$ and NaClO$_3$ were used to intercalate the 10 mesh graphite particles. using mechanical stirring. A cold-water bath was used for 10-15 min to run the reaction then the ambient temperature has been used. At first, H$_2$SO$_4$ was taken into a flask and slowly cooled down below 20 °C using a cold-water bath, then 2 g 10 mesh graphite particles were dispersed in H$_2$SO$_4$, and NaClO$_3$ was added to H$_2$SO$_4$ solution slowly. 300 ml cold deionized water was added to the mixture slowly at the end of the reaction time and then washed with deionized water. The filtered graphite particles were then dried at 60 °C for 24 h to obtain the intercalation compound GIC, named GIC-NaClO$_3$. To obtain the EG-NaClO$_3$ fillers, GICs were thermally expanded at 900 °C. 20-40 ml H$_2$SO$_4$ and 0.25-0.5 g NaClO$_3$ were utilized to optimize the amounts of reactants for the preparation of EG. Performed reaction conditions are mentioned in Table 2.

**Table 2** Chemicals and their compositions used for synthesis of GIC-NaClO$_3$ and EG-NaClO$_3$

| Sample name | 10 Mesh Graphite (g) | Amounts of reactants used | | Volume Ratio H$_2$SO$_4$: NaClO$_3$ | Time |
|---|---|---|---|---|---|
| | | H$_2$SO$_4$ (mL) | NaClO$_3$ (g) | | |
| GIC-NaClO$_3$ 1 & EG-NaClO$_3$ 1 | 2 | 20 | 0.25 | 200:1 | 30 min |
| GIC-NaClO$_3$ 2 & EG-NaClO$_3$ 2 | 2 | 20 | 0.25 | 200:1 | 1 h |
| GIC-NaClO$_3$ 3 & EG-NaClO$_3$ 3 | 2 | 20 | 0.5 | 100:1 | 30 min |



## 2.4 Preparation Method of Expanded Graphite-Polyetherimide (EG/PEI) Composite and Expanded Graphite (EG) Paper

Thermal expansion of graphite intercalated compound was carried out to obtain the worm-structured EG filler using a hot furnace at 900 °C. To achieve maximum expansion, the reacted GIC particles were kept inside the furnace for ~1 min. The solution casting technique was used to

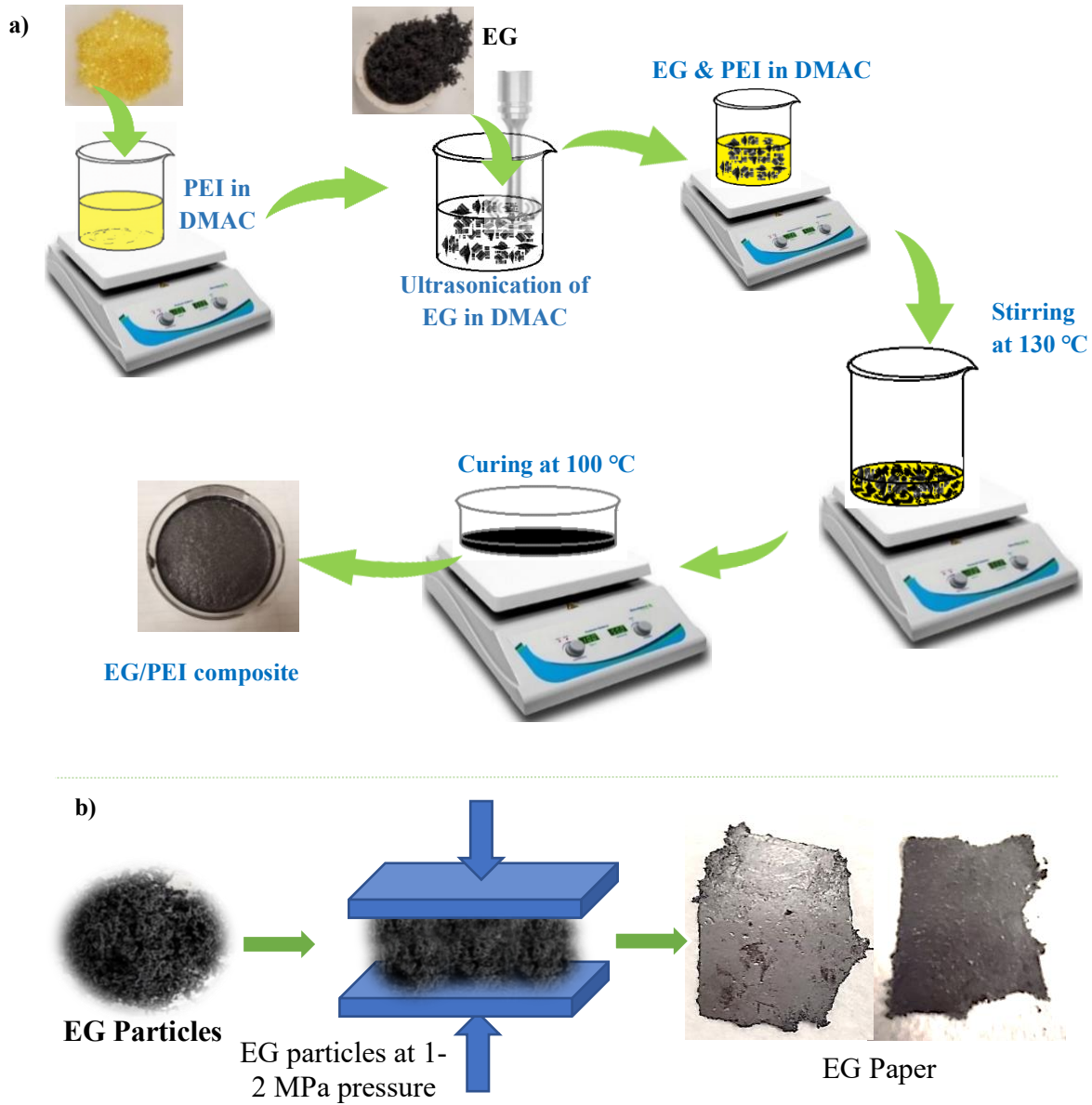

**Figure 2** a) Schematic preparation of EG/PEI composite, b) Preparation method of expanded graphite (EG) paper using compression molding. DMAC: N,N-dimethylacetamide



prepare the expanded graphite polymer composite, and polyetherimide (PEI) was chosen as the polymer matrix to mix with the EG filler. We have followed the fabrication procedure as shown in Figure 2a to prepare EG/PEI composite. GIC-$H_2O_2$ and GIC-$NaClO_3$ were separately expanded by using the thermal expansion technique to obtain the EG-$H_2O_2$ and EG-$NaClO_3$ particles. EG fillers were then dispersed into 20 mL DMAC. Separately PEI pellets were dissolved using 50 mL DMAC at 130 °C for 1 h. The DMAC solution with EG filler and dissolved polymer were mixed and blended for 3 h at 130 °C, followed by ultrasonication at 20% amplitude. Ultra-sonication time (1 min – 3 min) was optimized for the composite preparation process. Then, the mixture was cast into a glass petri dish and kept at 100 °C. The composite film was peeled off after 24-48 h. Composite films were prepared using different concentrations of EG filler- 2.5, 5, 7.5, and 10 wt%. The EG-$H_2O_2$/PEI and EG-$NaClO_3$/PEI composite films at 10 wt% filler loading were prepared using the same solution casting technique for different reaction conditions to compare the thermal conductivity value.

We also prepared the expanded graphite paper or EG paper using the compression molding technique on EG particles as illustrated in Figure 2b. Carver hot press was used to compress 2 g of EG particles together at room temperature under a pressure of 1-2 MPa and ~0.3 mm thick EG papers were fabricated. We prepared EG-$H_2O_2$ 1 and EG-$NaClO_3$ 1 papers separately, using EG-$H_2O_2$ 1 and EG-$NaClO_3$ 1 particles, respectively.

## 3. Characterization

The expansion volume (EV) was determined by exfoliating 1 g of GIC at a temperature of 900 °C for 30-60 s, and its volume was measured using a graduated cylinder. Then volume was recorded, and this datum was considered as EV value. To ensure accuracy, measurements were carried out in triplicate, and the average value was reported.

The Raman spectra was obtained using a Horiba Jobin-Yvon labRam HR instrument (HORIBA Scientific, France). Data were collected over the range from 3000 to 1000 cm$^{-1}$ using a laser wavelength $\lambda_L$ of 632 nm and a spectral resolution of 1.5 cm$^{-1}$. An Olympus BX 41 microscope with a 50× objective, a beam cross-sectional diameter of 25 μm, and 3 scans per sample was used to collect the spectra.



Rigaku SmartLab diffractometer (Rigaku Corporation, Japan) was used to produce the X-ray powder diffraction (XRD) patterns of GICs, and EGs at room temperature. A Cu Kα radiation (λ = 1.5406 Å) at 40 KV and 30 mA with a scan range of 5 to 80° and step size of 0.02° was used to collect the spectra. Bragg-Brentano configuration was used to collect the data at room temperature.

GICs and EGs were analyzed by Thermo Scientific K-alpha X-ray Photoelectron Spectroscopy (XPS) (ThermoFisher Scientific, Waltham, Massachusetts, USA), where Al Kα gun source was used to excite the sample, and measurement was carried out for acquisition time of ~68 s at 400 µm spot size. The passing energy of 200 eV was utilized to find the C, O and S peaks in this analysis spectrum. The elemental analysis of C, O & S and the abundance of functional groups were investigated using the Avantage software. To determine the functional group's peak position and atomic percentage, Avantage software was used for deconvoluting C1s curve fitting utilizing Gaussian and Lorentzian functions.

Morphological characterization of EG filler and EG/polymer composites was carried out by high-resolution Field Emission Environmental Scanning Electron Microscopy (Quattro S FE-ESEM, ThermoFisher Scientific, USA). This SEM was operated in secondary electron (SE) mode at an accelerating voltage of 20 kV.

## 4. Result & Discussion

### 4.1 Thermal Conductivity Measurement

A Netzsch LFA 467 Hyperflash was used to measure the through-thickness thermal diffusivity of the samples. The composite film samples of 0.5-0.8 mm thickness were cut into 12.5 mm diameter discs and coated with a thin layer of graphite paint. Thermal diffusivity measurements were performed at room temperature (23 °C) for 8-12 samples. Then, the thermal conductivity was calculated using $k = α×ρ×Cp$, where $k$, $ρ$, and $Cp$ represent the thermal conductivity, density, and specific heat constant of the sample, respectively. The density of composite sample was measured using the Pycnometer (Accu-Pyc II 1340, Micromeritics, US instrument). Rule of mixture formula



was used to calculate the specific heat of composite where the specific heat of pure polymer and expanded graphite were obtained from differential scanning calorimetry (DSC) (DSC 204F1 Phoenix, Netzsch, Selb, Germany). We measured the through-thickness thermal diffusivity (α) of EG paper using LFA 467 Hyperflash for 18-20 samples and averaged them to determine the thermal diffusivity value of EG paper.

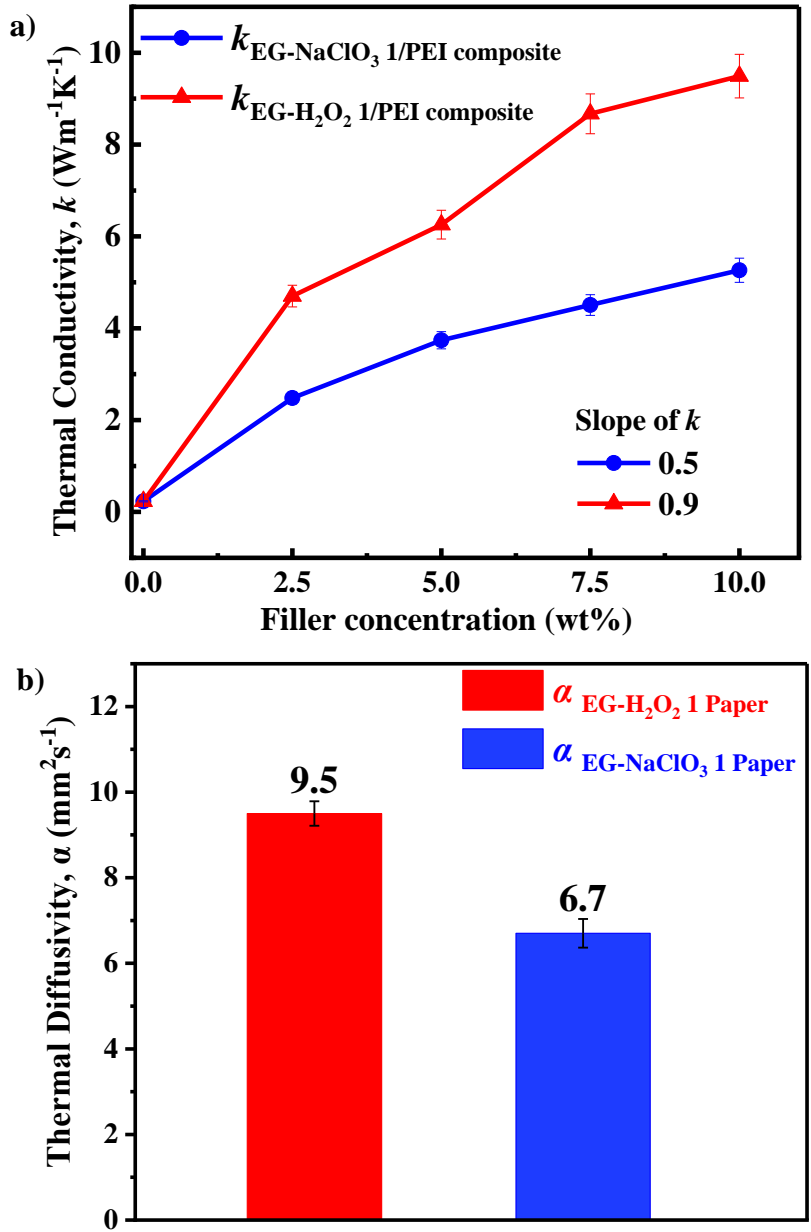

**Figure 3** a) $k$ value of EG-H$_2$O$_2$ 1/PEI & EG-NaClO$_3$ 1/PEI composite at 2.5, 5, 7.5 and 10 wt% filler concentration, b) $\alpha$ value of EG-H$_2$O$_2$ 1 & EG-NaClO$_3$ 1 paper.



## 4.2 Thermal Conductivity (*k*) Data Analysis

To compare the thermal properties of EG-$H_2O_2$ and EG-$NaClO_3$, the through thickness-thermal conductivity of expanded graphite polymer composite was measured. Comparison in thermal conductivity (*k*) value of EG/polymer composites for two auxiliary intercalating agents ($H_2O_2$ &

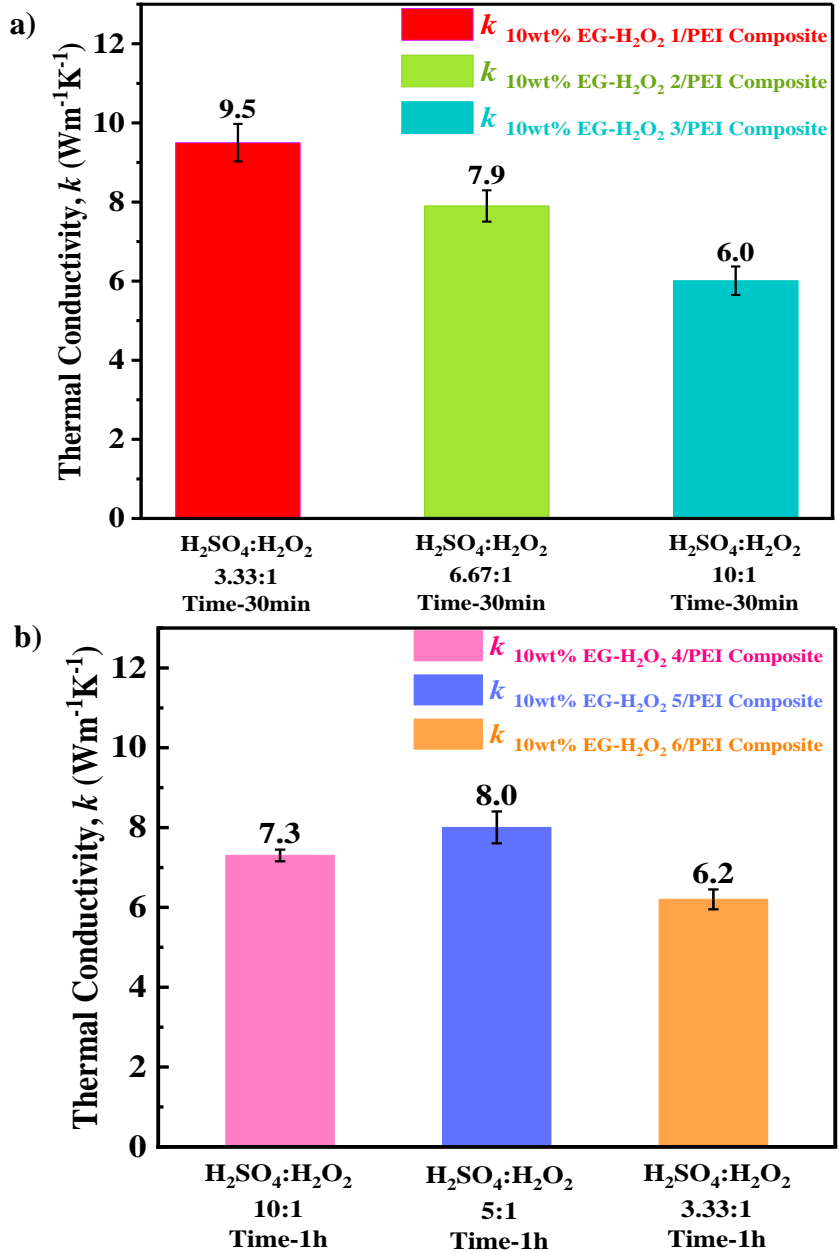

**Figure 4** *k* value of 10 wt% EG-$H_2O_2$/PEI composites for a) different volume ratio of $H_2SO_4$:$H_2O_2$ for 20, 40, and 60 mL of $H_2SO_4$ respectively, b) different volume ratio of $H_2SO_4$:$H_2O_2$ for 2, 4, and 6 mL of $H_2O_2$ respectively.



NaClO$_3$) at different weight percentages-2.5, 5, 7.5 & 10 wt% of EG filler content has been presented in Figure 3a. With the addition of 2.5 wt% filler, the $k$ of EG-H$_2$O$_2$ 1/PEI composites significantly increased to 4.7 Wm$^{-1}$K$^{-1}$ (representing an increase of ~1944% with respect to pure PEI), while the $k$ of EG-NaClO$_3$ 1/PEI composites only increased to 2.3 Wm$^{-1}$K$^{-1}$. Increasing the EG-H$_2$O$_2$ 1 filler concentration to 5 and 7.5 wt%, increased $k$ enhancement of EG-H$_2$O$_2$ 1/PEI composites by ~2620% and ~3670% respectively,. The slope of the EG-H$_2$O$_2$ /PEI composite's $k$ value (as shown in Figure 3a) is 0.9, on the other hand, the slope of the EG-NaClO$_3$ 1/PEI composite's $k$ is relatively low (0.5). $k$ value of EG-H$_2$O$_2$ 1/PEI composites reaches up to 9.5 Wm$^{-1}$K$^{-1}$ for 10 wt% EG-H$_2$O$_2$ 1 filler concentration, indicating a remarkable enhancement of ~4030% with respect to $k$ of pure PEI (0.23 Wm$^{-1}$K$^{-1}$ ). In comparison, a $k$ value of 3 Wm$^{-1}$K$^{-1}$ is achieved for EG-NaClO$_3$ 1/PEI composite with 10 wt% EG-NaClO$_3$ 1 filler concentration, showing ~2190% enhancement in $k$ value compared to pure PEI polymer. Such outstanding enhancement of $k$ for EG-H$_2$O$_2$ 1/PEI relative to EG-NaClO$_3$ 1/PEI composite reveals the superior effect of EG-H$_2$O$_2$ 1

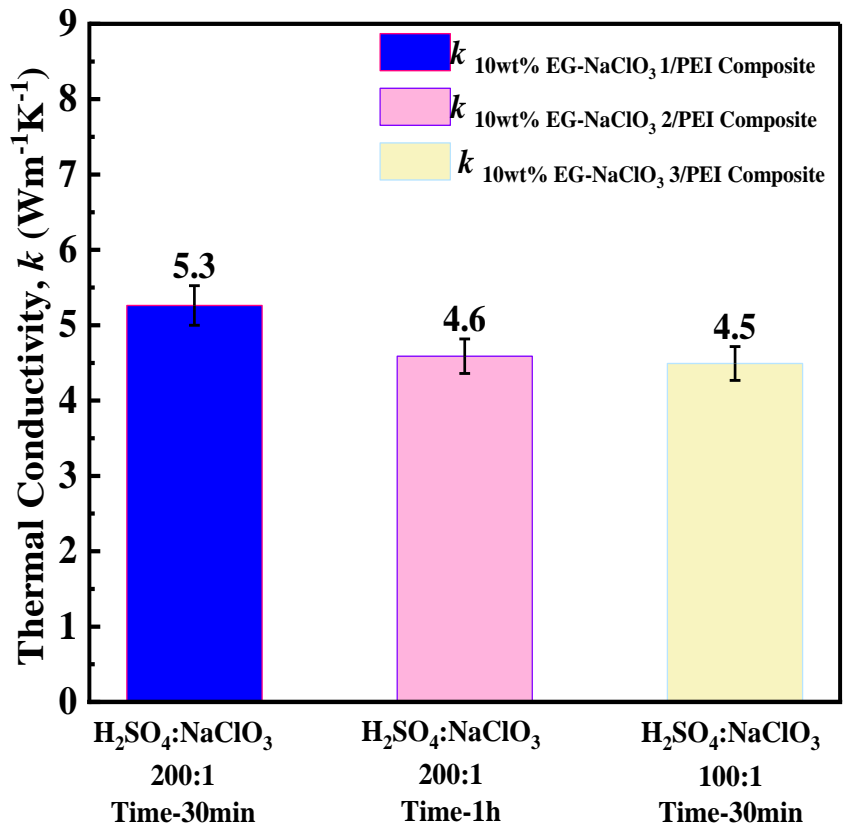

**Figure 5** $k$ value of 10 wt% EG-NaClO$_3$ /PEI composites with the different volume ratio of H$_2$SO$_4$: NaClO$_3$ and reaction time.



filler over the EG-NaClO$_3$ 1 filler. This superior effect of EG-H$_2$O$_2$ filler can be explained in terms of structural integrity, optimum expansion volume, and presence of higher C=C/C-C bonding. Most importantly, H$_2$O$_2$ leads to edge oxidation during the intercalation route I, whereas NaClO$_3$ leads to basal plane oxidation during the intercalation route II, as shown in XPS analysis (shown in section 4.5). To verify the thermal conductivity of the expanded graphite prepared from the two intercalation routes, EG papers have been prepared to measure the through-thickness thermal diffusivity ($\alpha$) and a higher $\alpha$ value has been found for EG-H$_2$O$_2$ 1 paper as shown in Figure 3b. ~42% enhancement in $\alpha$ value for EG-H$_2$O$_2$ 1 paper has been achieved compared to EG-NaClO$_3$ 1 filler.

We have performed multiple reactions to optimize the volume ratio of H$_2$SO$_4$:H$_2$O$_2$ and reaction time to achieve the highest thermal conductivity. Initially, keeping the amount of H$_2$O$_2$ the same, the quantity of H$_2$SO$_4$ was changed to 20 ml, 40 ml & 60 ml, and 10 wt% EG-H$_2$O$_2$/PEI composite composites were prepared for different volume ratios of H$_2$SO$_4$: H$_2$O$_2$ such as 3.33:1, 6.67:1 & 10:1. Figure 4a presents the $k$ value of 10 wt% EG-H$_2$O$_2$ /PEI composite for different volume ratios of H$_2$SO$_4$:H$_2$O$_2$, and shows a decrease in composite $k$ with an increase in H$_2$SO$_4$ quantity. This is due to strong intercalation with a higher amount of H$_2$SO$_4$, resulting in lower stage number in GIC-H$_2$O$_2$ and higher expansion volume of EG. The $k$ value of 10 wt% EG-H$_2$O$_2$ 1/PEI composite at a volume ratio of H$_2$SO$_4$:H$_2$O$_2$ (3:33:1) is found to be 9.5 Wm$^{-1}$K$^{-1}$; this decreases to 6 Wm$^{-1}$K$^{-1}$ for 10 wt% EG-H$_2$O$_2$ 3/PEI composite at a volume ratio of 10:1. Due to the strong intercalation, the defective structural morphology of EG filler is responsible for such reduction in $k$ value as shown through Raman analysis (section 4.4).

We have also observed a change in the $k$ value of the EG-H$_2$O$_2$/PEI composite with an increase in H$_2$O$_2$ quantity at a similar dosage of H$_2$SO$_4$. Figure 4b presents the thermal conductivity value of EG-H$_2$O$_2$/PEI composites for different volume ratios of 10:1, 5:1 & 3.33:1 at 2 ml, 4 ml, and 6 ml of H$_2$O$_2,$ respectively, and 1 h of intercalation time. $k$ value of EG-H$_2$O$_2$ 4/PEI composite is 7.3 Wm$^{-1}$K$^{-1}$ for 10 wt% EG-H$_2$O$_2$ 4 filler loading and $k$ value increases to 8 Wm$^{-1}$K$^{-1}$ for 10 wt% EG-H$_2$O$_2$ 5 filler loading. In contrast, 10 wt% EG-H$_2$O$_2$ 6 filler composition sample leads to a reduced $k$ value of 6.2 Wm$^{-1}$K$^{-1}$. For the case of 2 ml & 4 ml H$_2$O$_2$, the intercalation process does not complete with 20 ml H$_2$SO$_4$ for 60 min reaction time to reach optimum expansion volume. Optimum synthesis condition of 20 ml H$_2$SO$_4$, 6 ml H$_2$O$_2,$ and 30 min leads to optimum expansion volume and superior thermal conductivity of 9.5 Wm$^{-1}$K$^{-1}$ for 10 wt% EG-H$_2$O$_2$ 1/PEI composite.



We also have studied variation in *k* value of EG-NaClO$_3$/PEI composite with different quantities of H$_2$SO$_4$ & NaClO$_3$ and for different intercalation times. Figure 5 represents the *k* value of EG-NaClO$_3$/PEI composite for different volume ratios of 200:1 and 100:1. For a minimal amount of 0.25 g NaClO$_3$ and at a volume ratio of H$_2$SO$_4$: NaClO$_3$ (200:1), *k* value of EG-NaClO$_3$ 1/PEI composite reaches 3 Wm$^{-1}$K$^{-1}$ for 10 wt% EG-NaClO$_3$ 1 filler composition. The *k* value of EG-NaClO$_3$ 2/PEI composite is decreased to 4.6 Wm$^{-1}$K$^{-1}$ as the intercalation time increases from 30 min to 1 h. Higher intercalation time to prepare the GIC-NaClO$_3$ for the same amount of oxidant leads to more intercalation and lower stage number, leading to higher expansion volume. On the other hand, higher oxidation leads to higher structural defects as shown in XPS and Raman analysis, resulting in a lower *k* value. Also, a higher amount of NaClO$_3$ at volume ratio of H$_2$SO$_4$ : NaClO$_3$ (100:1) has a negative impact on EG-NaClO$_3$ 3/PEI composite's *k* value, leading to a lower *k* value of 4.5 Wm$^{-1}$K$^{-1}$ compared to EG-NaClO$_3$ 1/PEI composite. The measured *k* value does not increase at higher quantity of H$_2$SO$_4$ used in the reaction. Optimum reaction condition at a volume ratio of H$_2$SO$_4$: NaClO$_3$ (200:1) and 30 min intercalation time reveals a *k* value of 3 Wm$^{-1}$K$^{-1}$.

Additionally, the nature of oxidation greatly impacts the EG filler's structural integrity as well as the thermal conductivity of EG/PEI composite, as discussed in chapter 2 in detail. According to the XPS analysis, as discussed in section 4.5, we have found significant differences in oxidation degree and the presence of functional groups for GIC-H$_2$O$_2$ 1 & EG-H$_2$O$_2$ 1 in contrast to GIC-NaClO$_3$ 1 & EG-NaClO$_3$ 1.

Furthermore, the degree and location of oxidation significantly impact the structural integrity of the EG filler and the thermal conductivity of the EG/PEI composite in our recent work. We noticed a substantial difference in oxidation degree and the location-dependent functional groups for GIC-H$_2$O$_2$ 1 & EG-H$_2$O$_2$ 1 in comparison to GIC-NaClO$_3$ 1 & EG-NaClO$_3$ 1 in the XPS study reported in section 4. According to the location of functional groups of graphite oxide, the percentage of basal plane functional groups (C-O-C & C-O) are larger in quantity than edge functional groups (C=O, O=C-O) for NaClO$_3$. The GIC-H$_2$O$_2$ 1 & EG-H$_2$O$_2$ 1, on the other hand, contain a higher percentage of edge functional groups than basal plane functional groups.

It is important to mention that basal plane oxidation leads to distortion of sp$^2$ carbon structure due to attachment of the functional groups on the basal plane, increasing phonon scattering, and



dramatically reducing intrinsic thermal conductivity of graphene. On the contrary, edge oxidation avoids distortion of the basal plane area leading to higher intrinsic *k* of graphene.

### 4.3 Formation and Expansion Volume of GICs and EG Fillers

To intercalate graphite, $H_2O_2$ and $H_2SO_4$ were used to prepare GIC-$H_2O_2$, where $H_2SO_4$ was used as a primary intercalating agent and $H_2O_2$ as an auxiliary intercalating agent or oxidizing agent for the intercalation route I. The addition of oxidant, $H_2O_2$ into $H_2SO_4$ raises the temperature and causes self-decomposition of $H_2O_2$ according to Eq. For the case of intercalation using $H_2O_2$, the synthesis was carried out using a cold ice bath to maintain the temperature at 20-25 °C. We have utilized the strategy to intercalate graphite using oxidant $H_2O_2$ so that $H_2O_2$ helps to open up the edges of graphite with simultaneous insertion of $HSO_4^-$ and $SO_4^{2-}$ ions into graphite from $H_2SO_4$. Such intercalation process leads to a separation of graphene layers releasing $O_2$ (Eq. 6) and yielding GIC-$H_2O_2$ as shown in Figure 6a. Self-decomposition of $H_2O_2$ leads to a pre-expansion phenomenon before the GIC-$H_2O_2$ is exposed to high-temperature treatment for expansion.

The effect of different volume ratios of $H_2SO_4$: $H_2O_2$ on optimum expansion volume to achieve superior thermal conductivity is studied and described in the next section. The expansion volume of GIC and EG was recorded for different volume ratios of $H_2SO_4$: $H_2O_2$ in Table 3 as presented in Figure 7a-d and surface morphology through expansion was investigated using FE-ESEM in section 4.7. Intercalation process of graphite with an increased amount of $H_2SO_4$ produces more $O_2$ without allowing much time for this generated $O_2$ to escape, pushing the microchannels of graphite layers apart. Thus, an increased volume of graphite particles GIC-$H_2O_2$ is obtained after different reaction processes of 2 g 10 mesh graphite as shown in Figure 7a-d. The expansion volume after thermal shock increases with an increasing amount of $H_2O_2$ up to 3.33:1(20 ml:6 ml) volume ratio but the increased amount of $H_2O_2$ also produces a significant amount of $O_2$, which



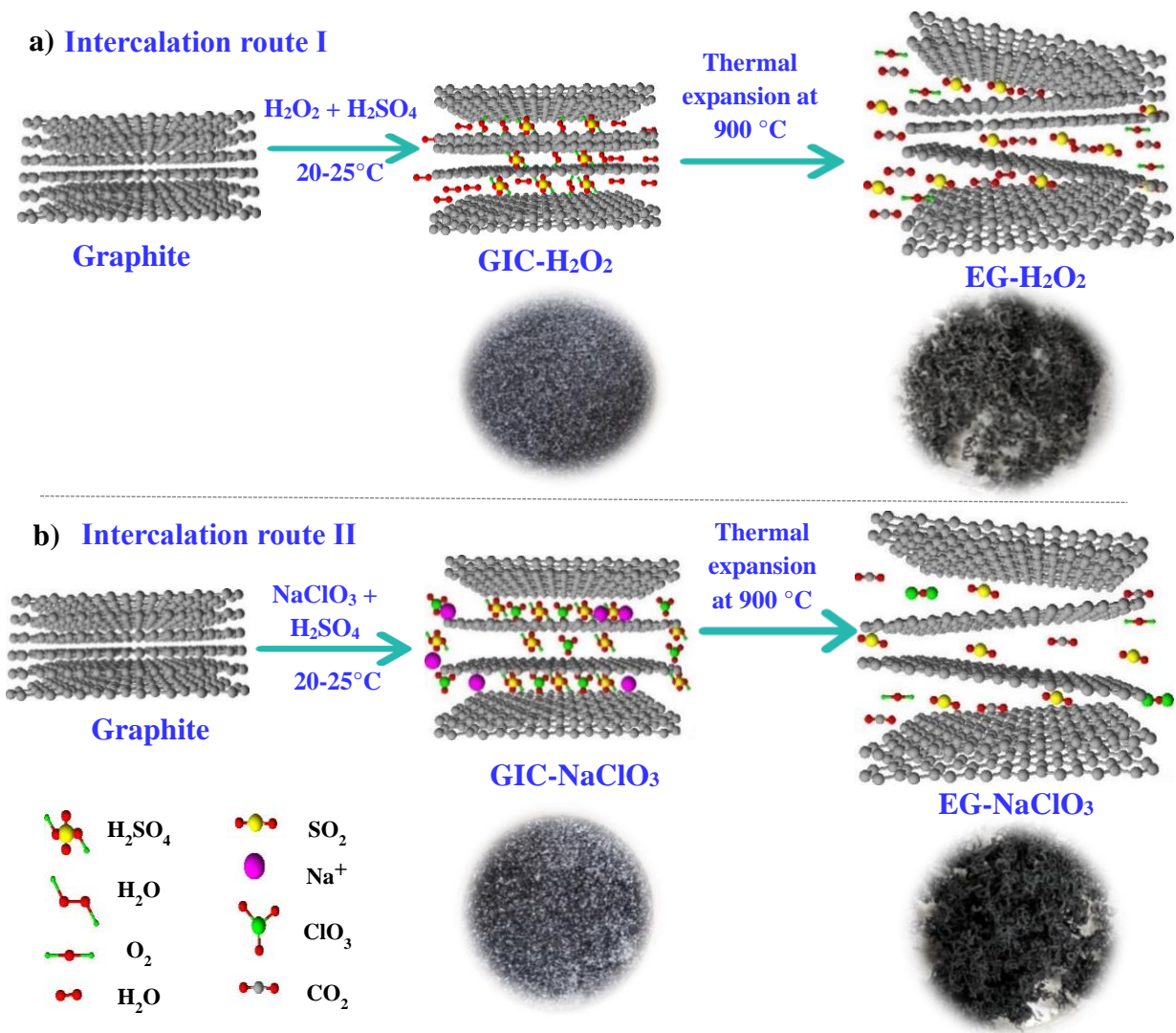

**Figure 6** a) Schematic illustration of the preparation process of a) GIC-H$_2$O$_2$ & EG-H$_2$O$_2$, b) GIC-NaClO$_3$ & EG-NaClO$_3$.

causes excessive exfoliation of graphite. Furthermore, higher amount of H$_2$SO$_4$ leads to more intercalation, following the reaction below, so the volume of graphite particles increases even after the reaction because of pre-expansion.

$$2H_2O_2 \rightarrow 2H_2 + O_2 \tag{5}$$

$$H_2SO_4 + H_2O_2 \rightarrow H_3O^+ + HSO_4^- + O\cdot \tag{6}$$

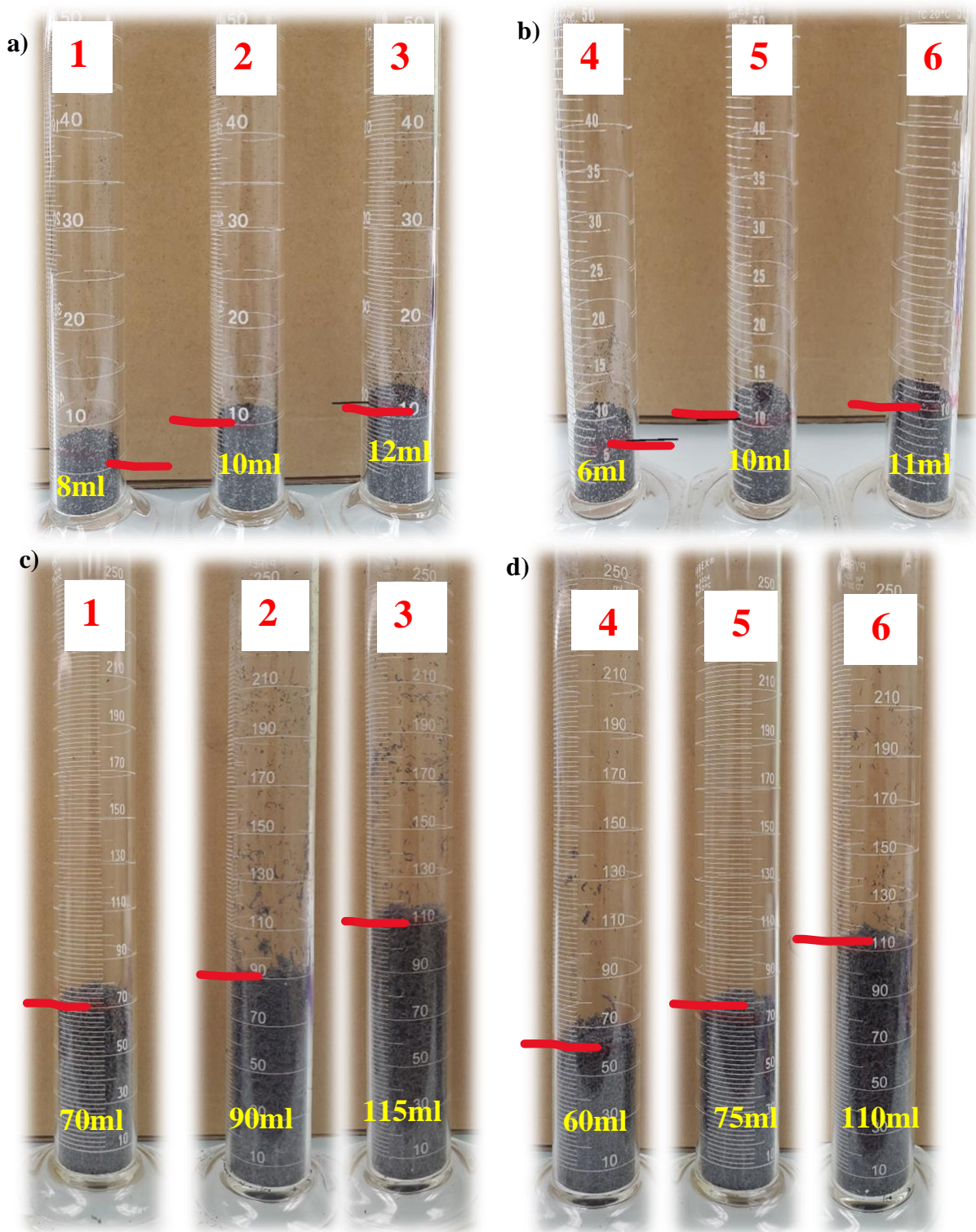

**Figure 7** Expansion volume (EV) of a) GIC-$H_2O_2$ 1, 2 & 3, b) GIC-$H_2O_2$ 4, 5 & 6, c) EG-$H_2O_2$ 1, 2 & 3, d) EG-$H_2O_2$ 4, 5 & 6.

Oxidant H$_2$O$_2$, known as a green oxidizer, allows oxidation to graphite with non-toxic bi-products and involves an easy washing procedure. Such oxidation process causes insignificant damage to graphite and provides optimum expansion volume to achieve higher thermal conductivity.

Another oxidant, sodium chlorate (NaClO$_3$) has been employed with H$_2$SO$_4$ to obtain sodium chlorate intercalated graphite compound (GIC-NaClO$_3$) and thermally expanded graphite, EG-NaClO$_3$ to observe the effect on thermal conductivity of EG-NaClO$_3$ polymer composite. The preparation process of GIC-NaClO$_3$ also involves performing reactions at room temperature of 20-25 °C. NaClO$_3$ has been used as a strong oxidizer for Brodie's and Staudenmaier's methods with other oxidizing agents[43,44]. With respect to previous work, the intercalation and oxidation effect of this oxidizer on EG has been observed in this work. With the intercalation of acceptor bisulfate compound, sodium ion gets diffused into the interlayer spacing as shown in Figure 6b and results in strong intercalation with lower stage number as evident through characterization analysis. During the reaction, oxygen groups attach to the basal plane structure and delamination possibly occurs. The reaction takes place during intercalation route II as mentioned below.

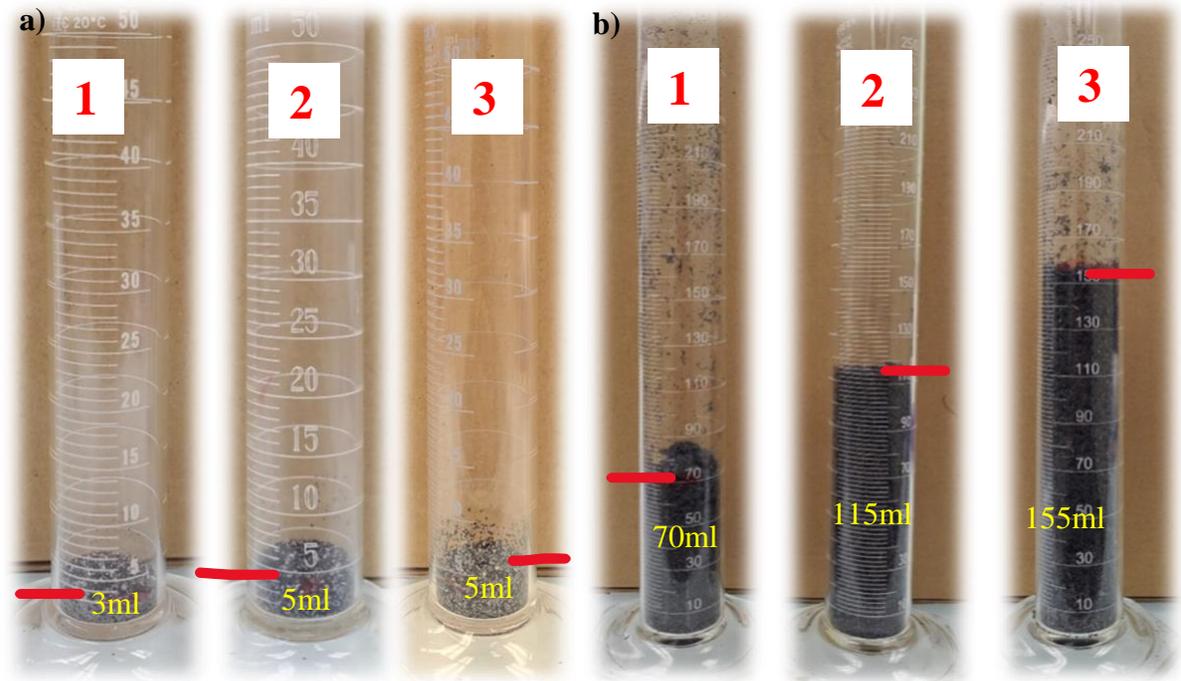

**Figure 8** Expansion volume of a) GIC-NaClO$_3$ 1, 2 & 3, and b) EG-NaClO$_3$ 1, 2 & 3.

$3NaClO_3 + 2H_2SO_4 \rightarrow 2NaHSO_4 + 2ClO_2 + NaClO_4 + H_2O$ (7)

**Table 3** Exfoliation volume and weight loss (%) for GICs & EG fillers.

| Sample name | Expansion volume (ml/g) | Weight loss (%) |
|:---:|:---:|:---:|
| GIC-$H_2O_2$ 1 | 8 | - |
| EG-$H_2O_2$ 1 | 70 | 17 |
| GIC-$H_2O_2$ 2 | 10 | - |
| EG-$H_2O_2$ 2 | 90 | 20 |
| GIC-$H_2O_2$ 3 | 12 | - |
| EG-$H_2O_2$ 3 | 115 | 42 |
| GIC-$H_2O_2$ 4 | 6 | - |
| EG-$H_2O_2$ 4 | 60 | 14 |
| GIC-$H_2O_2$ 5 | 10 | - |
| EG-$H_2O_2$ 5 | 75 | 24 |
| GIC-$H_2O_2$ 6 | 11 | - |
| EG-$H_2O_2$ 6 | 110 | 45 |
| GIC-$NaClO_3$ 1 | 3 | - |
| EG-$NaClO_3$ 1 | 70 | 21 |
| GIC-$NaClO_3$ 2 | 5 | - |
| EG-$NaClO_3$ 2 | 115 | 40 |
| GIC-$NaClO_3$ 3 | 5 | - |
| EG-$NaClO_3$ 3 | 155 | 55 |

Different volume ratios have been explored to achieve the optimum reaction condition for enhancing the thermal conductivity of EG-$NaClO_3$/PEI composites. The expansion volume of GICs and EGs with different volume ratios of $H_2SO_4$: $NaClO_3$ is presented in Figure 8a & b and recorded in Table 3. At a very small amount of $NaClO_3$, the expansion volume of EG is 70 ml/g indicating higher intercalation effect than EG-$H_2O_2$. Controlling the reaction process allows variation in the volume ratio. Optimum expansion volume of 70 ml/g of EG-$H_2O_2$ 1 and EG-$NaClO_3$ 1 has been utilized for preparing the composite to achieve the highest thermal conductivity in both the case.

## 4.4 Raman Analysis

Raman spectroscopy is used to investigate the structural and chemical changes of graphite structure because of the intercalation & oxidation process. Typically, the Raman spectra of graphite exhibit the characteristic G, D & 2D peaks at ~1572 cm$^{-1}$, ~1330 cm$^{-1}$, and ~2680 cm$^{-1}$ [45-47] respectively as presented in Figure 9a and Table 4 for 10 mesh graphite. The strong G peak is attributed to the vibration due to sp$^2$ hybridized carbon structure; the D peak is caused due to sp$^3$ hybridized carbon lattice defects[33]. The prominent 2D peak is the second order of the D peak caused by in-plane transverse optical phonons near the boundary of the Brillouin zone due to the electronic band structure, and always shows the double peak structure for graphite[48]. This shape and position of the 2D peak are useful for understanding the thickness of graphite[49].

**Table 4:** Raman peak positions and $I_D/I_G$ of GIC-$H_2O_2$ & EG-$H_2O_2$ fillers

| Sample name | G band (cm$^{-1}$) | D band (cm$^{-1}$) | 2D band (cm$^{-1}$) | $I_D/I_G$ |
|---|---|---|---|---|
| 10 Mesh Graphite | 1572 | 1332 | 2680 | 0.13 |
| GIC-$H_2O_2$ 1 | 1572 | 1326 | 2677 | 0.05 |
| EG-$H_2O_2$ 1 | 1573 | 1331 | 2680 | 0.04 |
| GIC-$H_2O_2$ 2 | 1571 | 1324 | 2676 | 0.06 |
| EG-$H_2O_2$ 2 | 1572 | 1324 | 2679 | 0.05 |
| GIC-$H_2O_2$ 3 | 1571 | 1331 | 2680 | 0.09 |
| EG-$H_2O_2$ 3 | 1582 | 1335 | 2687 | 0.1 |

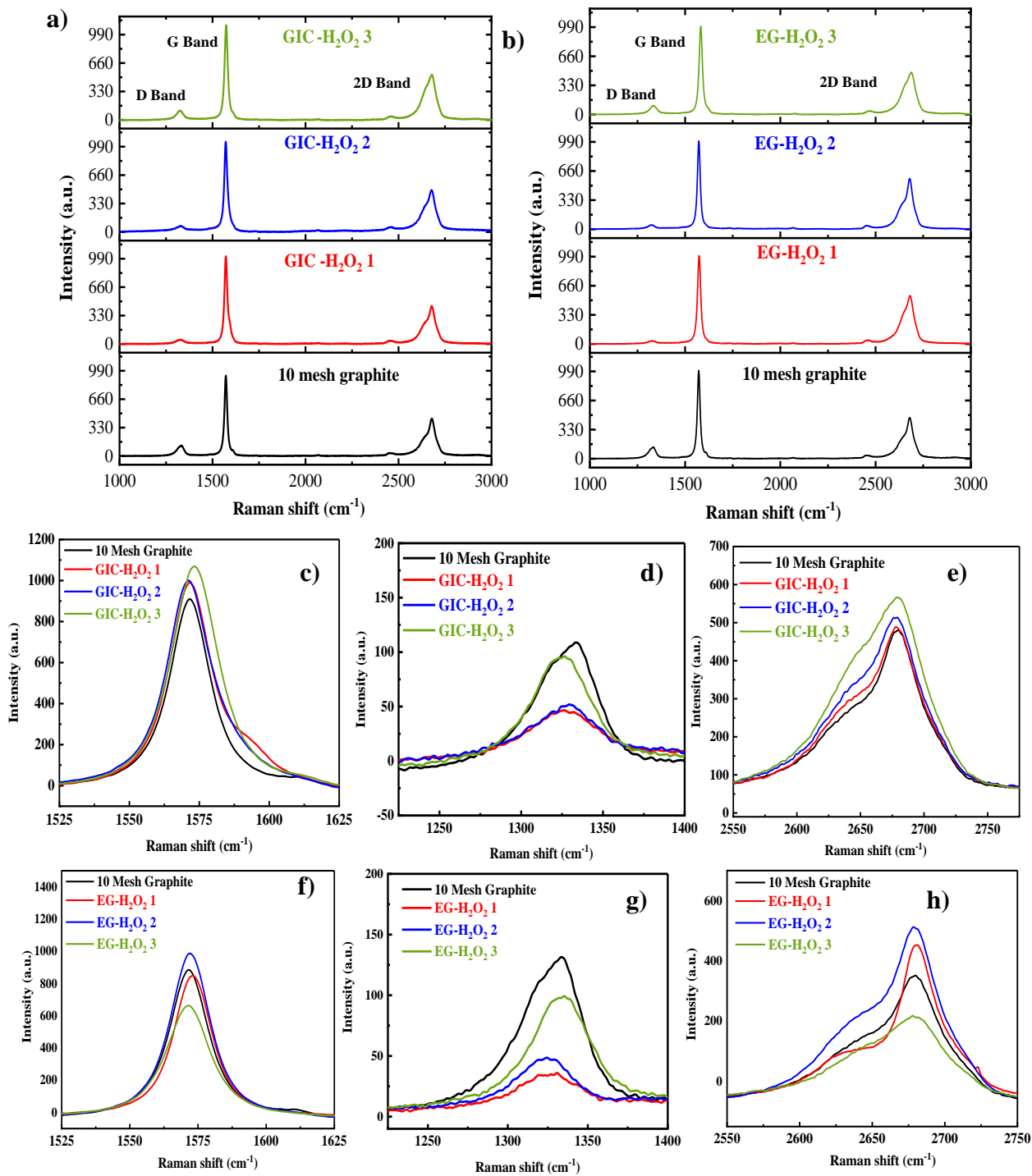

**Figure 9** Raman spectra of a) GIC-H$_2$O$_2$ 1, 2 & 3, and b) EG-H$_2$O$_2$ 1, 2 & 3. c) G band line, d) D band line, e) 2D band line of GIC-H$_2$O$_2$ 1, 2 & 3, f) G band line, d) D band line, e) 2D band line of EG-H$_2$O$_2$ 1, 2 & 3.

The peak positions for GICs and EG fillers for intercalating agent $H_2O_2$ are presented in Table 4. We have estimated the $I_D/I_G$ ratio for the GICs and EG fillers after the intercalation and thermal expansion to analyze the defective state of structure in graphite layers as reported in Table 4. $I_D/I_G$ of 10 mesh graphite indicates the presence of defects in graphite structure as associated with the functional groups attached to the graphite shown in XPS analysis. Figure 9a & b present the Raman spectra of GIC-$H_2O_2$ 1, 2 & 3 and EG-$H_2O_2$ 1, 2 & 3 respectively to compare the change in carbon material due to the intercalation process and heat treatment. Figure 9 c, d & e show the G, D & 2D peak intensity of GIC-$H_2O_2$ 1, 2 & 3 with an increase in the amount of $H_2SO_4$ respectively, and Figure 9 f, g & h show the G, D & 2D peak intensity of EG-$H_2O_2$ 1, 2 & 3 after the thermal treatment. Figure 9 c-h clearly depict the change in intensity with the increase of $H_2SO_4$ for intercalation with $H_2O_2$. $I_D/I_G$ ratios for GIC-$H_2O_2$ 1, 2 & 3 are 0.05, 0.06, and 0.09, indicating an increase in degree of defect with the higher amount of $H_2SO_4$. Due to self-decomposition of $H_2O_2$, GICs have even lower $I_D/I_G$ ratio than 10 mesh graphite. Similarly, $I_D/I_G$ ratios for EG-$H_2O_2$ filler show similar trend in the degree of defect. $I_D/I_G$ ratio increases from 0.04 to 0.1 with the change in the amount of $H_2SO_4$ from 20 ml to 40 ml. Additionally, the change in peak position and asymmetric broader peak for GICs and EG fillers compared to 10mesh graphite as shown in Figure 9 c-h refers to the oxidation effect. In Figure 9 e and h, the 2D peak position does not degrade to much lower wavenumber compared to 2D peak position of 10mesh graphite indicating higher stage number for the intercalation route with $H_2O_2$. The stronger G peak intensity in EG-$H_2O_2$ 1, 2 & 3 fillers refers to the better structural integrity, consistent with the XRD (Section 4.6) and XPS analysis (Section 4.5). The thermal conductivity value of EG-$H_2O_2$ 1, 2 & 3 PEI composites appears to be similarly decreased with higher degree of defect in graphite's structure.

**Table 5** Raman peak positions and $I_D/I_G$ of GIC-$NaClO_3$ & EG-$NaClO_3$ fillers

| Sample name | G band (cm$^{-1}$) | D band (cm$^{-1}$) | 2D band (cm$^{-1}$) | $I_D/I_G$ |
|---|---|---|---|---|
| 10 Mesh Graphite | 1572 | 1332 | 2680 | 0.13 |
| GIC-$NaClO_3$1 | 1574 | 1326 | 2680 | 0.35 |
| EG-$NaClO_3$ 1 | 1574 | 1332 | 2676 | 0.25 |
| GIC-$NaClO_3$ 2 | 1578 | 1325 | 2667 | 0.75 |
| EG-$NaClO_3$ 2 | 1575 | 1332 | 2678 | 0.28 |
| GIC-$NaClO_3$ 3 | 1577 | 1332 | 2678 | 0.62 |
| EG-$NaClO_3$ 3 | 1573 | 1324 | 2680 | 0.26 |

The peak positions and $I_D/I_G$ ratio for GICs and EG fillers for intercalating agent $NaClO_3$ are presented in Table  Figure 10a & b portray the Raman spectra of GIC-$NaClO_3$ 1, 2 & 3 and EG-$NaClO_3$ 1, 2 & 3 indicating the change in the defective state of carbon material due to the intercalation process and heat treatment respectively. Figure 10 c, d & e show the G, D & 2D peak intensity of GIC-$NaClO_3$ 1, 2 & 3 with the increase in the quantity of $NaClO_3$ and intercalation time. Figure 10 f, g & h show the G, D & 2D peak intensity of EG-$NaClO_3$ 1, 2 & 3 after the thermal treatment compared to pristine graphite. Figure 10 c-h clearly illustrate the change in intensity with the increase of $H_2SO_4$ for intercalation with $NaClO_3$. $I_D/I_G$ ratio increases from 0.35 to 0.75 for GIC-$NaClO_3$ 1 and 2, indicating an increase in the degree of defect with the higher intercalation time. An increase in $I_D/I_G$ ratio (0.62) for GIC-$NaClO_3$ 3 signifies an increased degree of defect with the higher amount of $NaClO_3$. $I_D/I_G$ ratios of EG-$NaClO_3$ 1, 2 & 3 are 0.25, 28 & 0.23, show the similar trend in degree of defect. Overall, the $I_D/I_G$ ratio for EG-$NaClO_3$ is higher due to the effect of stronger intercalating agent, $NaClO_3$. Figure 10 c & f show the broader G peak and change in peak position and Figure 10 d & g show the broader D peak of GIC-$NaClO_3$ 1, 2 & 3 and EG-$NaClO_3$ 1, 2 & 3 because of higher disorder in the graphite structure. Also, another peak ~1600 $cm^{-1}$ close to the G peak (shown in Figure 10 d & g) suggests large-scale damage and deformation to the graphitic crystalline structure due to oxidation [refer]. The comparison of $I_D/I_G$ ratios of EG-$NaClO_3$ 2 and 3 reveals that the intercalation time affects more than the amount of $NaClO_3$. Therefore, the higher thermal conductivity value of EG-$NaClO_3$ 1 PEI composite than other EG-$NaClO_3$/PEI composite is due to such a lower $I_D/I_G$ ratio.

Raman spectra of GIC-$H_2O_2$ 1 & GIC-$NaClO_3$ 1 and EG-$H_2O_2$ 1 & EG-$NaClO_3$ 1 fillers for intercalating agent $H_2O_2$ and $NaClO_3$ are depicted in Figure 11a &b to observe graphitic structural damage, happened during the intercalation and expansion process. We can see $I_D/I_G$ ratio (0.25) of EG $NaClO_3$ 1 is higher than $I_D/I_G$ ratio (0.04) for EG-$H_2O_2$ 1. Figure 11 c, d & e show the G, D & 2D peak intensity of EG-$H_2O_2$ 1 & EG-$NaClO_3$ 1. The D peak of Figure 11 d & f shows the higher intensity of EG-$NaClO_3$ 1 than EG-$H_2O_2$ 1 filler.  Another peak, ~1610 $cm^{-1}$ close to G band, is visible in Figure 11c & f for GIC-$NaClO_3$ and EG-$NaClO_3$, which can be explained as the induced defect due to the strong oxidation effect is absent in GIC-$H_2O_2$ & EG-$H_2O_2$ spectrum. A broader 2D peak of EG-$NaClO_3$ 1 suggests a lower stage number than EG-$H_2O_2$ 1 Figure 11 e & h. The structural integrity of carbon material plays a key role in the thermal properties of carbon-based polymer composites field[50,51]. Incorporation of EG-H$_2$O$_2$ 1 filler with lower I$_D$/I$_G$ ratio results in superior thermal conductivity due to less defective carbon structure.

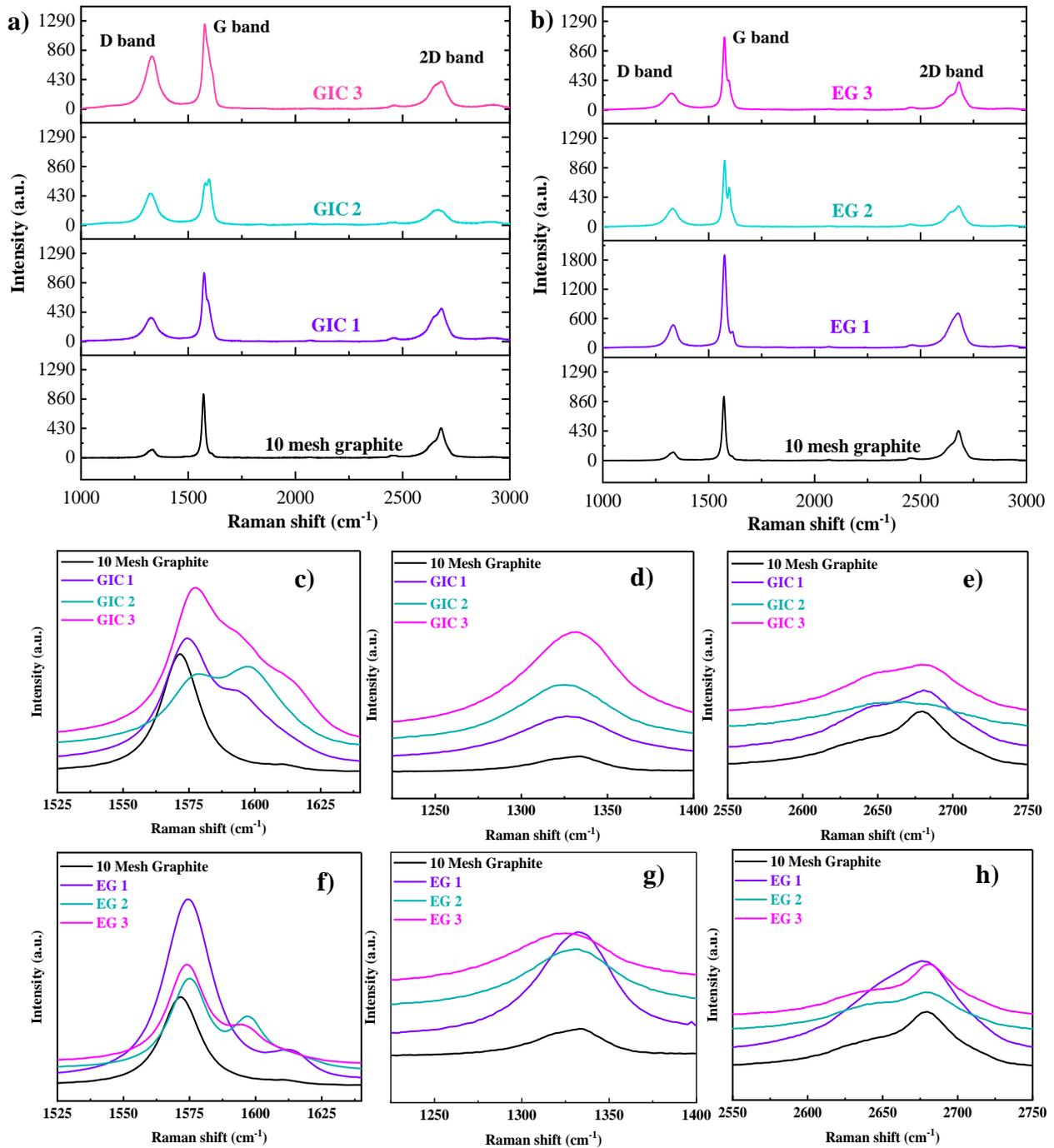

**Figure 10** Raman spectra of a) GIC-NaClO$_3$ 1, 2 & 3, and b) EG-NaClO$_3$ 1, 2 & 3. c) G band line, d) D band line, and e) 2D band line of GIC-NaClO$_3$ 1, 2 & 3, f) G band line, g) D band line, and h) 2D band line of EG-NaClO$_3$ 1, 2 & 3.

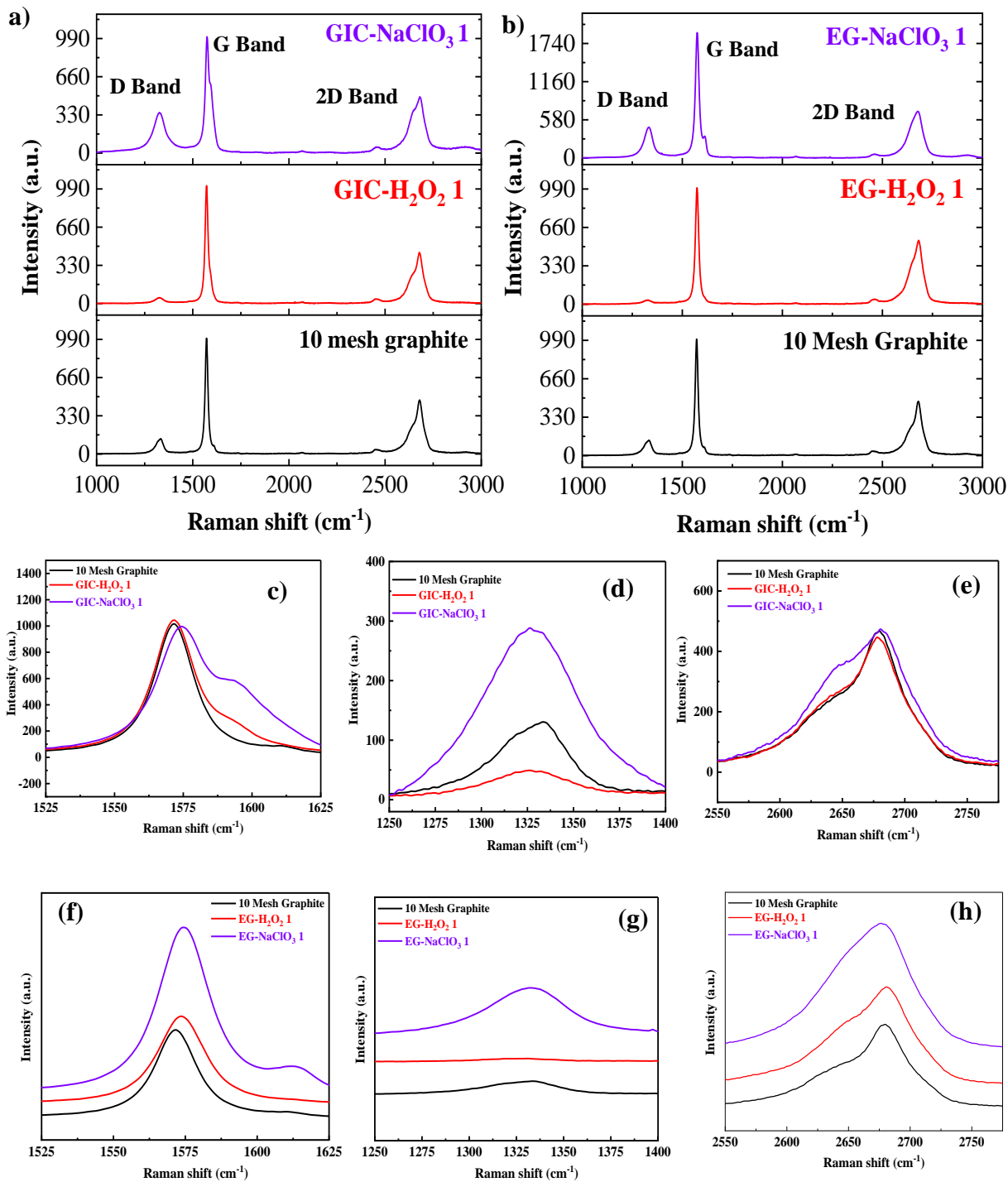

**Figure 11** Raman spectra of a) GIC-$H_2O_2$ 1 & GIC-$NaClO_3$ 1, and b) EG-$H_2O_2$ 1 & EG-$NaClO_3$ 1. c) G band line, d) D band line, and e) 2D band line of GIC-$H_2O_2$ 1 & GIC-$NaClO_3$ 1, f) G band line, g) D band line, and h) 2D band line of EG-$H_2O_2$ 1 & EG-$NaClO_3$ 1.

## 4.5 Chemical Composition Analysis of GICs and EG Fillers by XPS

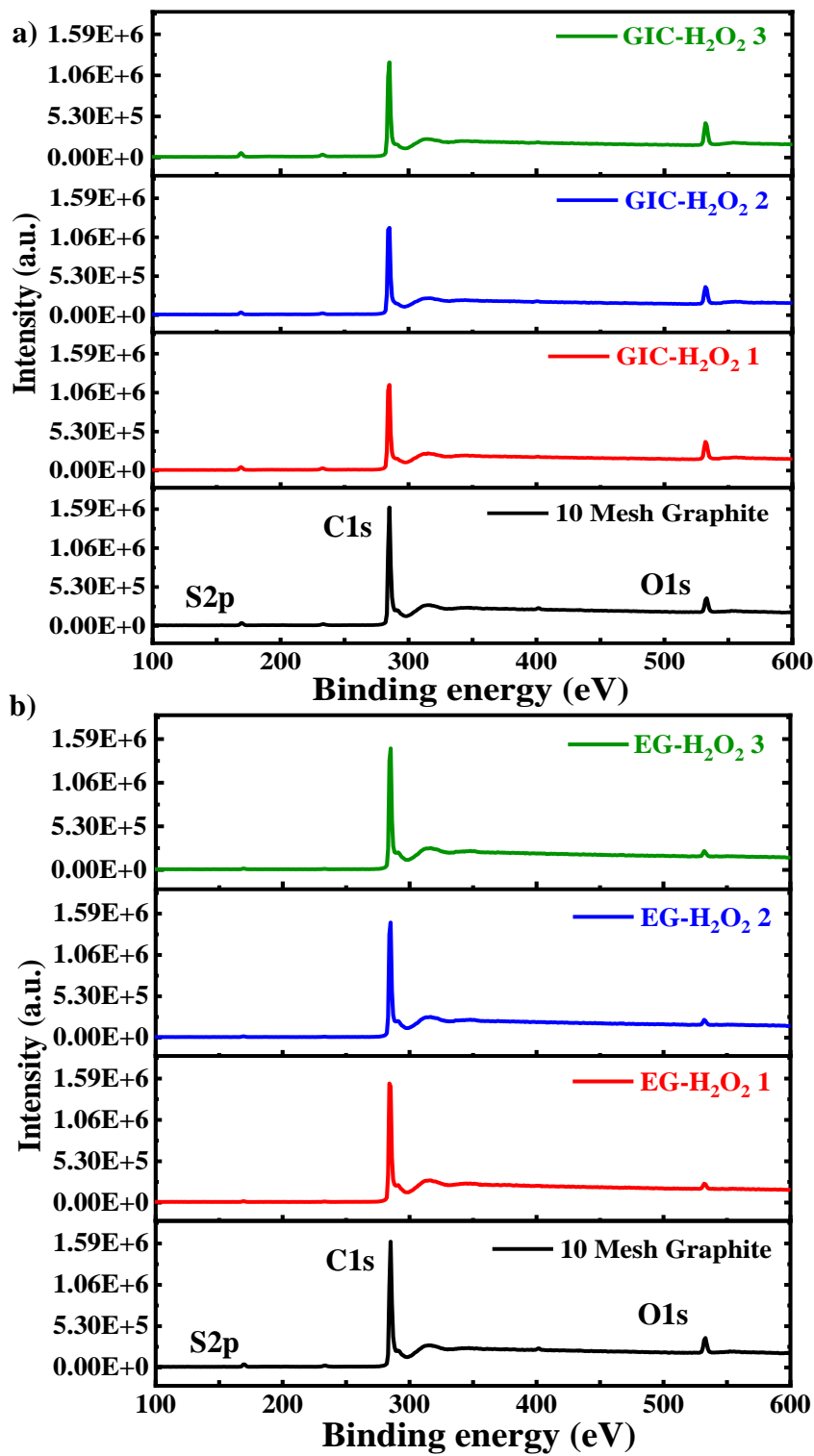

**Figure 12** a) XPS spectra of GIC-H$_2$O$_2$ 1, 2 & 3 and b) EG-H$_2$O$_2$ 1, 2 & 3.

X-ray photoelectron spectroscopy (XPS) analysis reveals the chemical composition and binding states of GICs and EG fillers. Figure 12a & b and Figure 13a &b show the presence of C1s peak at ~285 eV, strong O1s peaks at ~532 eV, and weaker S2p peak at ~169 eV[33,52] for 10 mesh graphite, GICs, and EG fillers because of intercalation using $H_2O_2$ & $NaClO_3$ and thermal expansion. Table 6 presents the atomic percentage of C, O, and S, and C/O ratio for GICs and EG fillers. The atomic percentage of C shows that the C content decreases and O increases with higher intercalation using a larger quantity of $H_2SO_4$. The increase in $H_2SO_4$:$H_2O_2$ volume ratio from 3.33 to 10 results in a reduction in C/O ratio from 10.8 for GIC-$H_2O_2$ 1 to 8.8 for GIC-$H_2O_2$ 3. After thermal expansion, the C/O ratio similarly decreases from 44.2 for EG-$H_2O_2$ 1 to 23 for EG-$H_2O_2$ 3. We have achieved higher $k$ value (9.5 $Wm^{-1}K^{-1}$) for EG-$H_2O_2$ 1 composition, which decreases to 6 $Wm^{-1}K^{-1}$ for EG-$H_2O_2$ 3 due to higher oxidation.

Figure 13 a & b present the XPS spectra of GICs and EG fillers, respectively showing three characteristic peaks of C, O, and S for intercalation route, $NaClO_3$. Analysis of the C/O ratio of GIC 1, 2 & 3 suggests a decrease in carbon content with the use of the higher quantity of strong oxidant, $NaClO_3$ and higher intercalation time, which also raises the content of sulfur. For optimum thermal conductivity, the C/O ratio is found to be 6.7 at a lower amount of $NaClO_3$ and lower intercalation time of 30 min. After thermal treatment, the C/O ratio of EG shows a similar trend as for GIC. EG-$NaClO_3$1 has the highest carbon content of almost 95% and provides a higher $k$ value relative to other EG-$NaClO_3$ fillers. In addition, sulfur content still is present in EG-$NaClO_3$ fillers due to the substantial intercalation effect. C/O ratio of EG-$H_2O_2$ 1 (44.2) is significantly higher than that of EG-$NaClO_3$ 1 (22.7), indicating that higher carbon content is preserved in intercalation route I after thermal expansion, resulting in superior $k$ value of EG-$H_2O_2$ 1/PEI composite than that of EG-$NaClO_3$1/PEI composite.

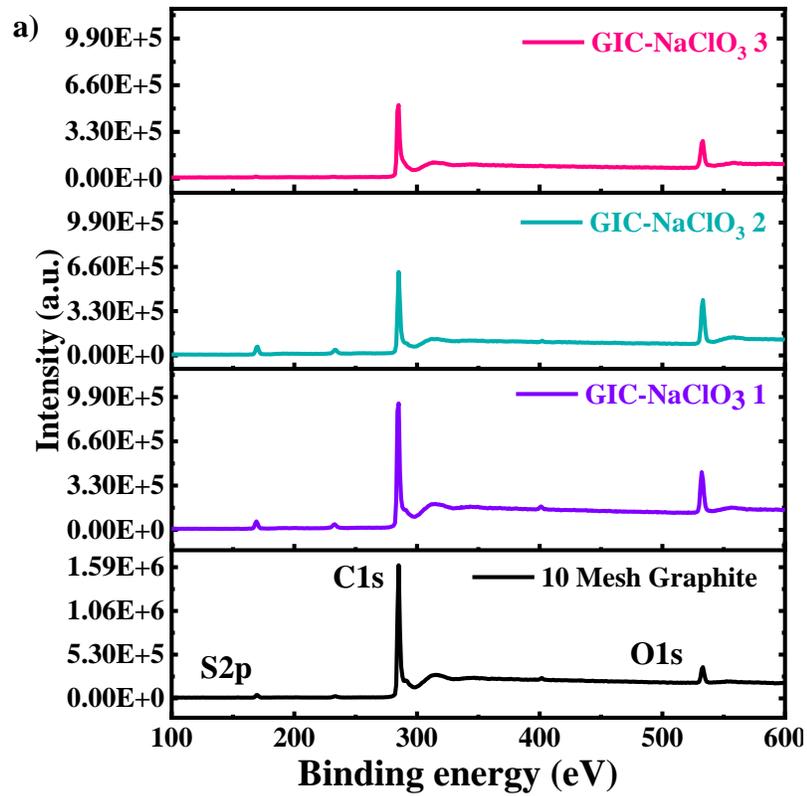

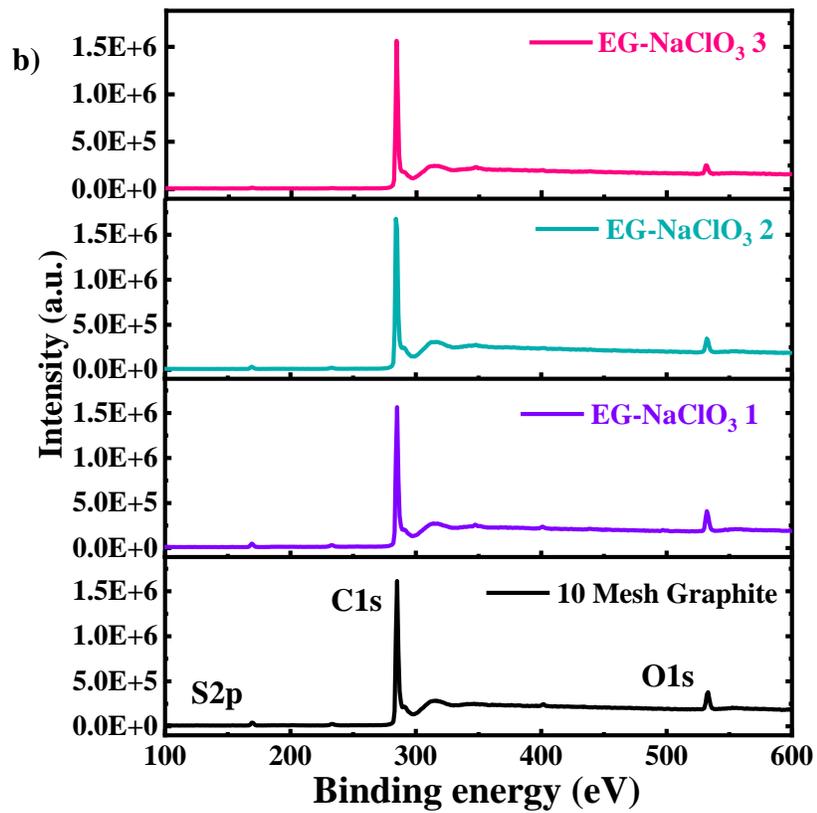

**Figure 13** a) XPS spectra of GIC-NaClO$_3$ 1, 2 & 3, and b) EG-NaClO$_3$ 1, 2 & 3.

**Table 6** Atomic composition and C/O ratio by XPS analysis of graphite, GICs, and EG- fillers

| Samples | Atomic Composition by XPS (at%) | | | C/O ratio |
|---|---|---|---|---|
| | C (~285 eV) | O (~532 eV) | S (~169 eV) | |
| **10 mesh graphite** | 93.44 | 47 | 1.09 | 17.1 |
| GIC-$H_2O_2$ 1 | 90.4 | 8.35 | 1.24 | 10.8 |
| EG-$H_2O_2$ 1 | 97.79 | 2.21 | - | 44.2 |
| GIC-$H_2O_2$ 2 | 89.47 | 8.81 | 1.72 | 10.1 |
| EG-$H_2O_2$ 2 | 96.5 | 3.5 | - | 27.6 |
| GIC-$H_2O_2$ 3 | 87.84 | 9.99 | 2.17 | 8.8 |
| EG-$H_2O_2$ 3 | 96.2 | 3.8 | - | 23 |
| GIC-$NaClO_3$ 1 | 84.85 | 12.63 | 2.52 | 6.7 |
| EG-$NaClO_3$ 1 | 94.97 | 4.18 | 0.85 | 22.7 |
| GIC-$NaClO_3$ 2 | 78 | 19.94 | 4.26 | 3.8 |
| EG-$NaClO_3$ 2 | 91.85 | 6.81 | 1.34 | 13.5 |
| GIC-$NaClO_3$ 3 | 80.73 | 16.59 | 2.68 | 4.9 |
| EG-$NaClO_3$ 3 | 92.52 | 7.09 | 0.4 | 18.6 |

We have also shown the comparison in atomic precentage of functional groups between the GICs and EG fillers for $H_2O_2$ and $NaClO_3$ at their optimum condition in Table 7. To further analyze, the high-resolution carbon spectra of GIC-$H_2O_2$ 1, GIC-$NaClO_3$ 1, EG-$H_2O_2$ 1, and EG-$NaClO_3$ 1 are shown in Figure 14a-d. Sharper peaks are visible for both the GIC and EG filler case for $H_2O_2$ intercalation compared to $NaClO_3$. To understand the differences in functional groups achieved through the two intercalation routes ($H_2O_2$ and $NaClO_3$), the C1s high-resolution XPS spectra were further analyzed and resolved by curve fitting. The deconvoluted spectra of GIC-$H_2O_2$ 1 (shown in Figure 14a) reveal the presence of C-C/C=C graphitic carbon (284.84 eV), the hydroxyl/epoxide group (~286 eV), and the carbonyl/carboxyl group (~288 eV), respectively[53,54]. Basal plane functional groups (hydroxyl/epoxide groups) are present in higher quantities in GIC-$NaClO_3$ 1 and EG-$NaClO_3$ 1 (Figure 14b & d), due to the fact that $NaClO_3$ mostly attacks the basal plane of the graphite structure . In comparison, deconvoluted XPS spectra of GIC-$H_2O_2$ 1 & EG-$H_2O_2$ 1 show that the edge functional groups are more prominent as seen in Figure 14 a and c. It is noticeable that EG-H$_2$O$_2$ 1 displays a stronger intensity peak of graphitic carbon (C-C/C=C) compared to EG-NaClO$_3$ 1; this directly correlates to better structural integrity of expanded graphite prepared from oxidant H$_2$O$_2$. The atomic percentage of epoxy and hydroxyl functional groups (C-O-C/C-OH) as well as the edge functional groups (C=O/O=C-OH) (as shown in Table 7) reveal higher degree of edge functionalization in the case of EG-H$_2$O$_2$ 1. The presence of higher degree of edge functional groups relative to basal plane functional groups in graphite's structure limits the structural damage in EG-H$_2$O$_2$ 1 sample while presence of higher degree of basal plane functional groups lead to higher damage for NaClO$_3$ intercalation. The scheme of intercalation with $H_2O_2$, thus drastically preserves the basal plane structure which leads to superior thermal transport in EG-$H_2O_2$ 1 sample.

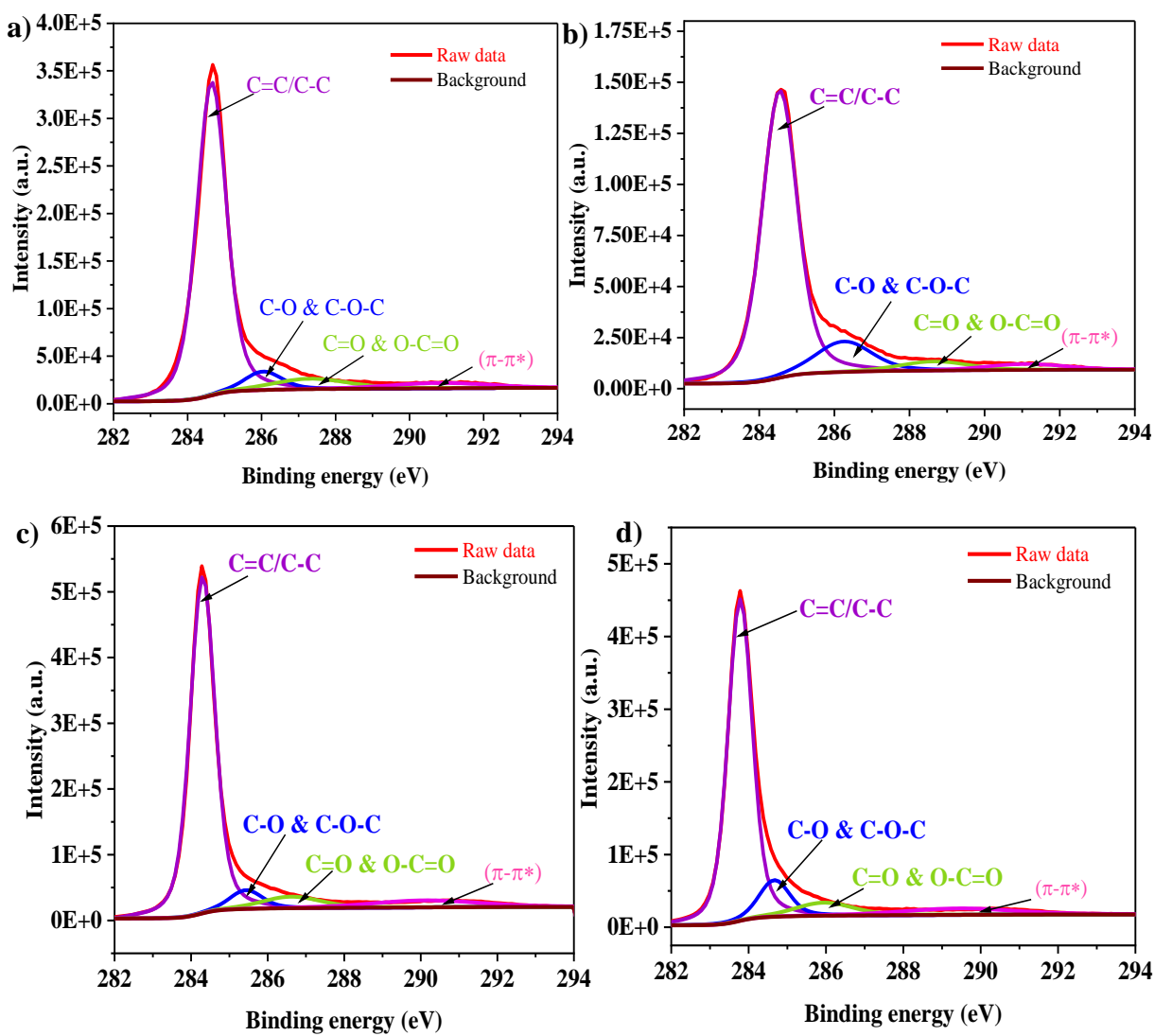

**Figure 14** C1s core level XPS spectra for a) GIC-H$_2$O$_2$ 1, b) GIC-NaClO$_3$ 1, c) EG-H$_2$O$_2$ 1, and d) EG-NaClO$_3$ 1.

**Table 7** Relative atomic percent composition of functional groups of GICs & EG fillers

| Sample name | C-C/C=C | C-OH & C-O-C | C=O & O=C-OH |
|:---:|:---:|:---:|:---:|
| GIC-$H_2O_2$ 1 | 82.4 | 91 | 7.16 |
| GIC-$NaClO_3$ 1 | 711 | 13.54 | 6.22 |
| EG-$H_2O_2$ 1 | 80.31 | 9 | 6.56 |
| EG-$NaClO_3$ 1 | 73.8 | 10.78 | 8.75 |

## 4.6 Analysis of Interlayer Spacing of GICS and EG Fillers through XRD

XRD analyzes the structural characteristics of the obtained GICs and EG fillers. XRD allows us to understand oxidation and intercalation effects on GICs and EG filler through the two auxiliary intercalating agents, $H_2O_2$ and $NaClO_3$. For the 10-mesh graphite, the sharp diffraction peak at $2\theta = 26.3°$ corresponds to an interlayer spacing of 3.38 Å. This peak position is present at a lower angle than $2\theta = 26.6°$ for pristine graphite[55-58], corresponding to the standard 3.34 Å interlayer spacing of the (002) crystal phase for untreated graphite. Table 8 represents the peak position and d-spacing values of graphite GICs and EG fillers for both the intercalation routes.

**Table 8** $2\theta$ and d-spacing values of graphite, GICs, and EG fillers using XRD analysis

| Sample name | 10 Mesh Graphite | GIC-$H_2O_2$ 1 | GIC-$H_2O_2$ 2 | GIC-$H_2O_2$ 3 |
|---|---|---|---|---|
| $2\theta$ | 26.3 | 25, 26.4 | 22, 24 | 23, 26.3 |
| d-spacing (Å) | 3.38 | 3.49, 3.37 | 3.53, 3.51 | 3.52, 3.38 |
| **Sample name** | **10 Mesh Graphite** | **EG-$H_2O_2$ 1** | **EG-$H_2O_2$ 2** | **EG-$H_2O_2$ 3** |
| $2\theta$ | 26.3 | 26.5 | 26.5 | 26.5 |
| d-spacing (Å) | 3.38 | 3.36 | 3.36 | 3.36 |
| **Sample name** | **10 Mesh Graphite** | **GIC-$NaClO_3$ 1** | **GIC-$NaClO_3$ 2** | **GIC-$NaClO_3$ 3** |
| $2\theta$ | 26.3 | 28 | 24.7 | 26 |
| d-spacing (Å) | 3.38 | 3.45 | 3.6 | 3.48 |
| **Sample name** | **10 Mesh Graphite** | **EG-$NaClO_3$ 1** | **EG-$NaClO_3$ 2** | **EG-$NaClO_3$ 3** |
| $2\theta$ | 26.3 | 26.4 | 26.3 | 26.4 |
| d-spacing (Å) | 3.38 | 3.37 | 3.38 | 3.37 |

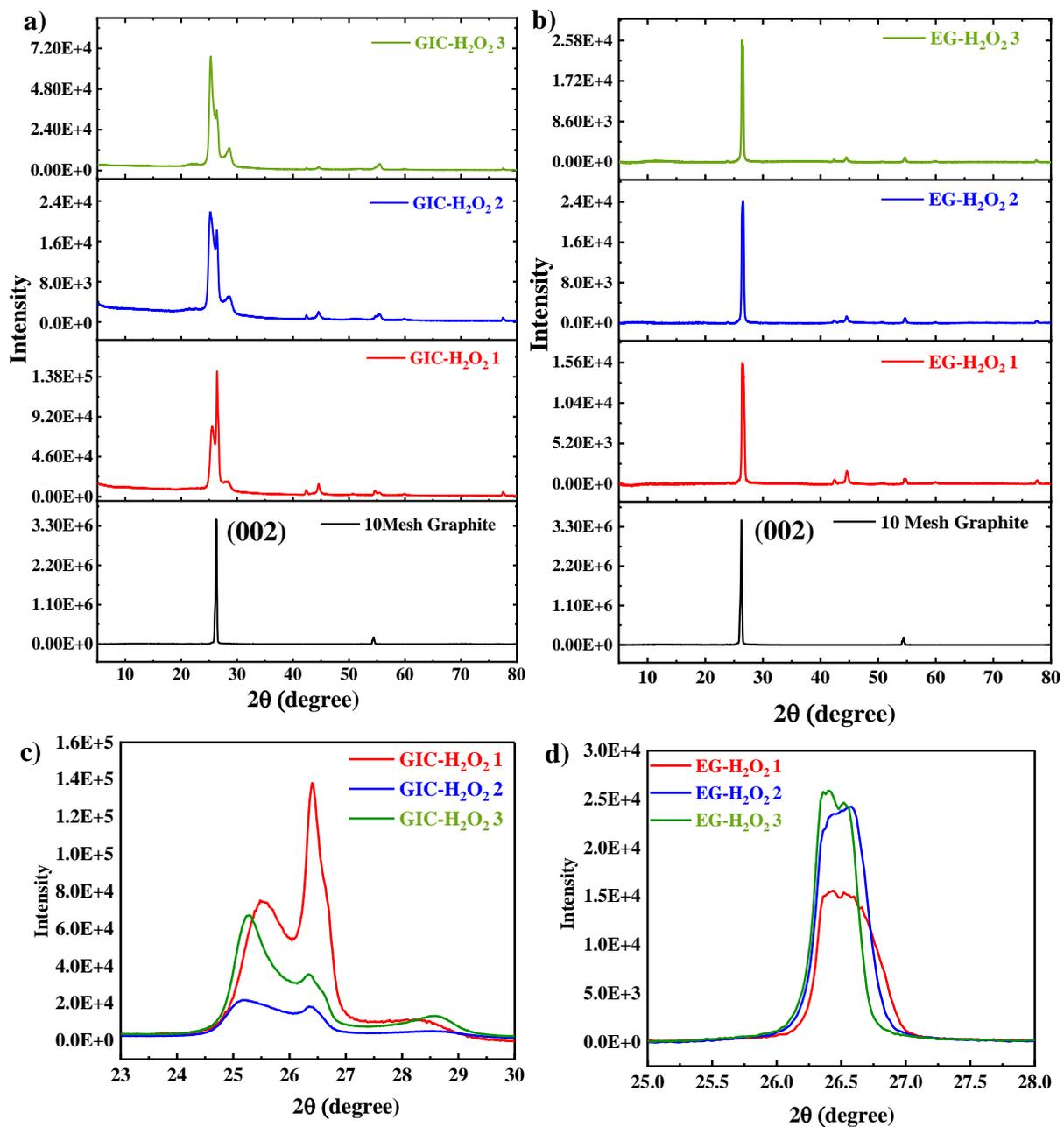

**Figure 15** XRD spectra of a) GIC-$H_2O_2$ 1, 2 & 3, b) EG-$H_2O_2$ 1, 2 & 3; Inset of XRD spectra of c) GIC-$H_2O_2$ 1, 2 & 3, and d) EG-$H_2O_2$ 1, 2 & 3.

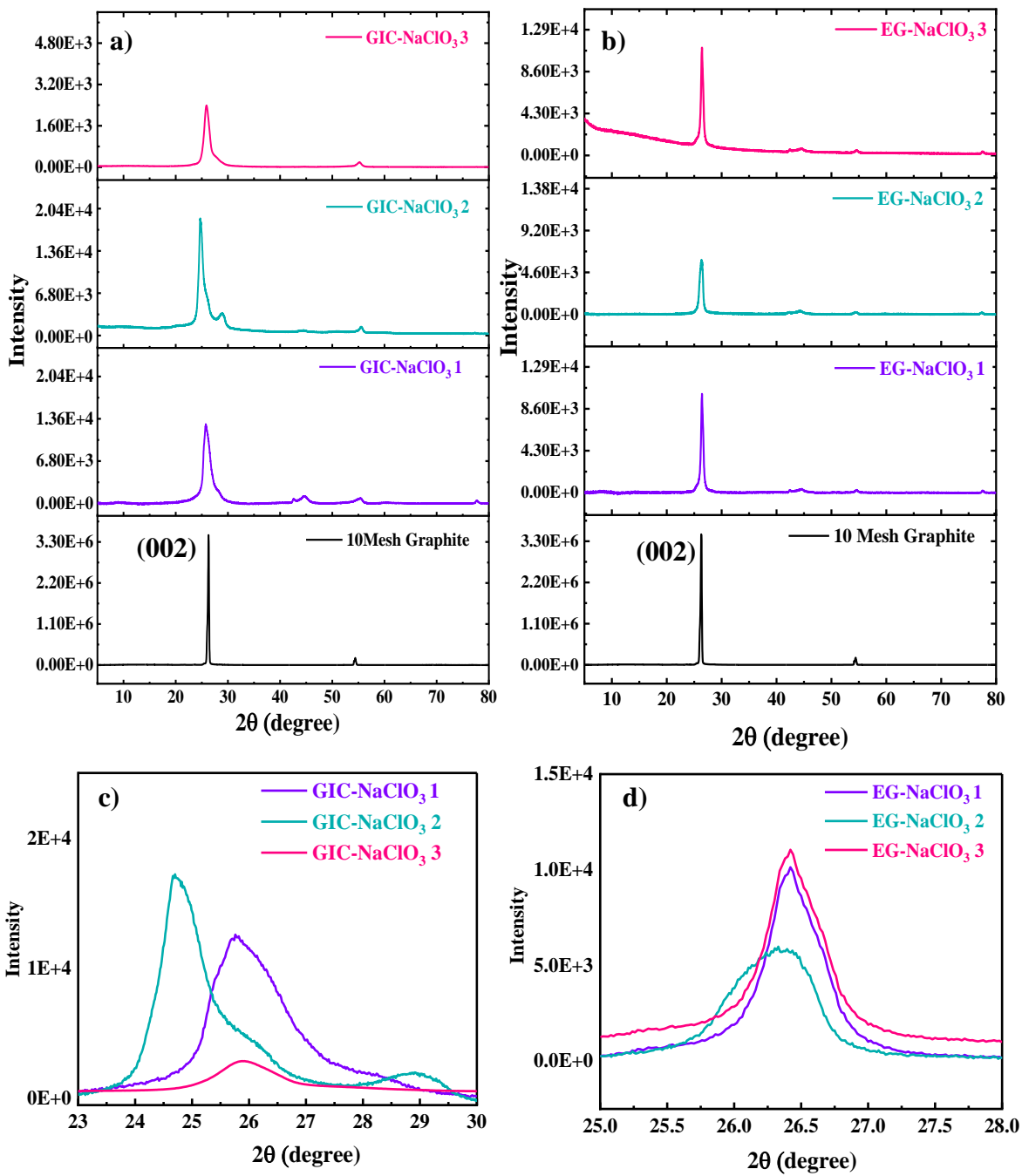

**Figure 16** XRD spectra of a) GIC-NaClO₃ 1, 2 & 3, b) EG-NaClO₃ 1, 2 & 3; Inset of XRD spectra of c) GIC-NaClO₃ 1, 2 & 3 and d) EG-NaClO₃ 1, 2 & 3.

Figure 15 a & b illustrate the diffraction peaks of GIC-H$_2$O$_2$ and EG-H$_2$O$_2$, respectively, and demonstrate the XRD pattern before and after the thermal treatment for an increasing amount of H$_2$SO$_4$. The only signal observable in the XRD pattern of natural graphite is (002), which shows reflections in the perpendicular direction (c-axis) of the hexagonal planes of graphite. Compared to 10 mesh graphite, the (002) peaks in the GIC-$H_2O_2$ 1, 2 & 3, are noticeably widened. The 002 peak is seen at 2θ angle (around ~26.4°) for GIC-$H_2O_2$ 1 and intensity becomes weaker (as exhibited in Figure 15c) with the increased amount of intercalating agents due to the disorder in

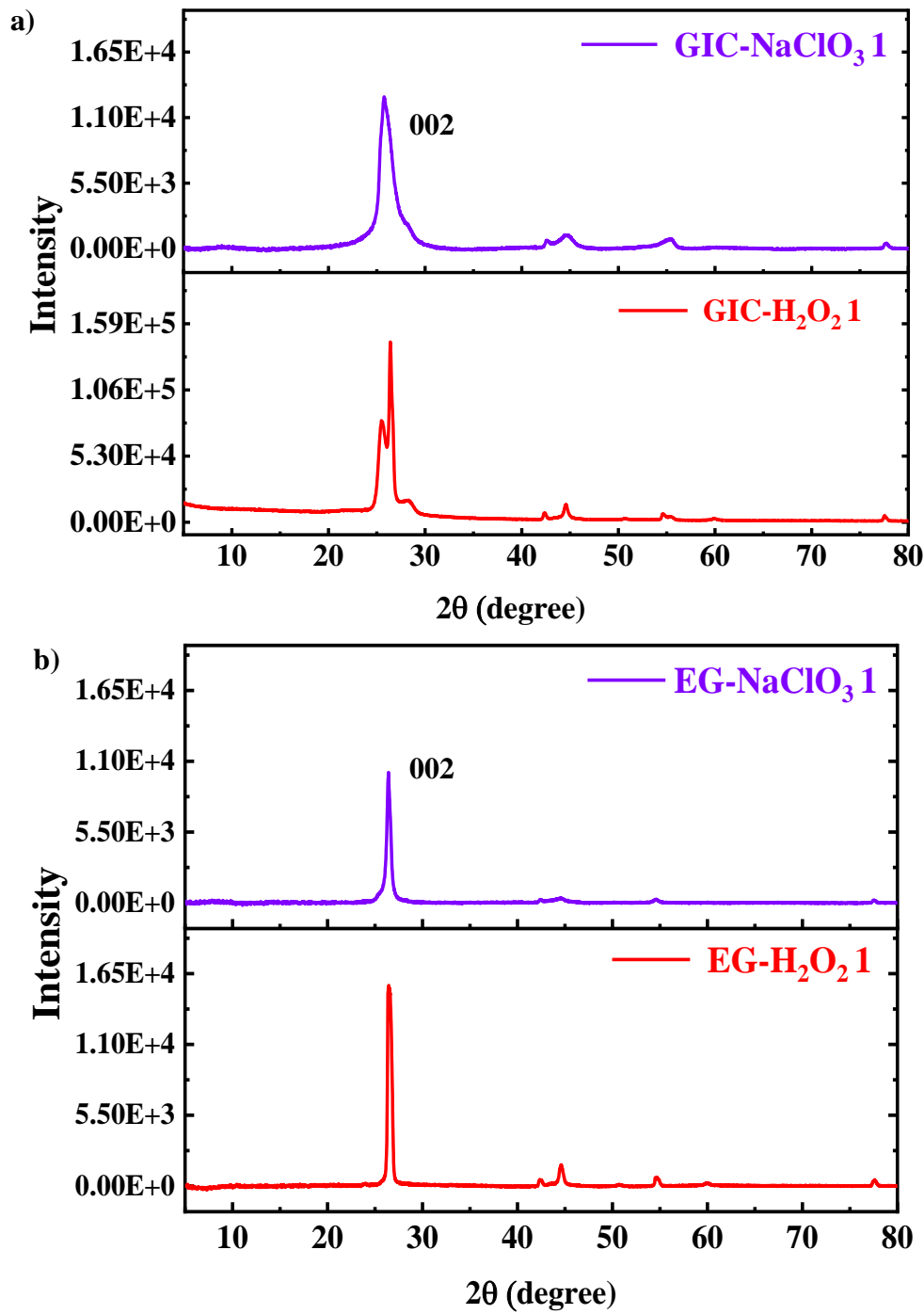

**Figure 17** XRD spectra of a) GIC-$H_2O_2$ 1 & GIC-$NaClO_3$ 1, b) EG-$H_2O_2$ 1 & EG-$NaClO_3$ 1.

graphitic morphology[59] after the pre-expansion process. Also, the double peak nature in GIC-$H_2O_2$

1, 2 & 3 is attributed to the presence of intercalated compounds. Interlayer spacing critically becomes higher with the intercalation effect as indicated by a clear diminution in intensity of (002) peaks (visible in Figure 15c). After thermal expansion, the intensity becomes even weaker than the GICs. Comparatively stronger peaks of EG-H$_2$O$_2$ 1, 2 & 3 at 2θ = 25° are regained (as shown in Figure 15b) as the intercalated compounds evaporate in gaseous form during thermal treatment. Such aligned peak position of EG fillers also indicates the existence of intact chemical structure of graphite and ordered morphology[35,60]. The interlayer spacing between graphene layers is 3.36 Å which is slightly higher than pristine graphite layers. In addition, the (002) peak intensity is still wider (Figure 15d) than the untreated graphite suggesting the expanded nature of EG filler. This interconnected and stacked structure of EG enables better thermal transport throughout the polymer composite[61].

In contrast, the peak shape and position of GICs and EG filler obtained from NaClO$_3$ intercalation route are displayed in Figure 16 a &b. The reflection peaks of GIC-NaClO$_3$ 1, 2 & 3 at 2θ = 28°, 24.7°, and 26°, lead to interlayer spacings of 3.45, 3.6, and 3.48 Å, respectively. The asymmetric but intense peak shape of GIC-NaClO$_3$ 2 shifts to a lower 2θ angle, reflecting the insertion of many intercalating species. Higher intercalation time causes insertion of higher intercalated compounds between the layers and maximum peak shift as observed for GIC-NaClO$_3$ 2 (Figure 16c). After expanding the GICs at 900 °C, the EG-NaClO$_3$ 1, 2 & 3 are subjected to the change in peak position at 2θ = 26.4, 26.5 and 26.5 as discussed earlier, resulting in the interlayer spacing of 3.37, 3.38 and 3.37 Å respectively.

We have compared the XRD spectra of GICs and EG fillers, obtained at optimum conditions for the two intercalation routes in Figure 17 a & b. It is noticeable that the interlayer spacing is higher for the case of EG-NaClO$_3$ 1 than EG-H$_2$O$_2$ 1 filler and the broader peak in EG-NaClO$_3$ 1 compared to EG-H$_2$O$_2$ 1 (Figure 17 b) reveals the effect of higher intercalation, and lower stage number in EG-NaClO$_3$ 1. Also the broader peak in EG-NaClO$_3$ 1 is caused due to the reduced crystalline structure and an increase in amorphous regions[62]. The larger change in peak position for GIC-NaClO$_3$ is attributed to the presence of higher oxygen moieties in NaClO$_3$ case, resulting in lower thermal conductivity of EG-NaClO$_3$ 1. On the other hand, stronger intensity accompanied with only a slight change in peak position of EG-H$_2$O$_2$ 1 filler is attributed to lower disorder in the crystal structure of graphite layers[63] leading to higher thermal conductivity of EG-H$_2$O$_2$ 1 filler.

## 4.7 Structure and Morphology Characterization of GICs and EG Filler

The morphology and porosity of the graphite structure due to intercalation and worm-like thermally expanded graphite have been studied using FE-ESEM. The FE-ESEM image of 10 mesh graphite particles with an average lateral size of 800 μm is shown in Figure 18. Figure 19-22. illustrate the structural change in GICs, resulting from chemical intercalation of 10 mesh graphite using $H_2O_2$ and $NaClO_3$ separately and thermally treated expanded graphite, observed by FE-ESEM.

Figure 19 a, b & c show images of GIC-$H_2O_2$ 1 highlighting the expansion phenomenon as generated from $H_2O_2$ intercalation reaction. Magnified images of GIC-$H_2O_2$ 1 (Figure 19 b) show a highly porous structure with sharp edges that emerge due to the reaction scheme between $H_2SO_4$ and $H_2O_2$ producing a large amount of $O_2$ that tries to escape from the graphite layers. The measured average lateral size of GICs and thickness of the graphite walls are ~500 μm and ~1 μm respectively. The intercalated chemicals are decomposed by heating GICs to 900 °C, resulting in puffed up materials (Figure 19 d & h) with a very porous structure (as shown in Figure 19e-i).

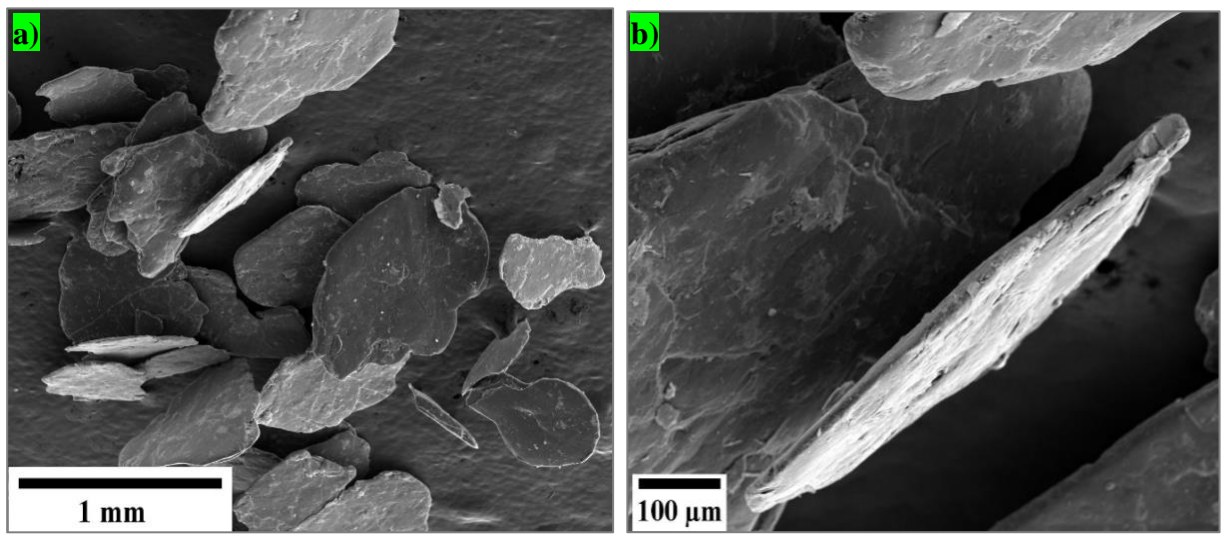

**Figure 18** FE-ESEM images of 10 mesh graphite a) ×35, b) ×150.

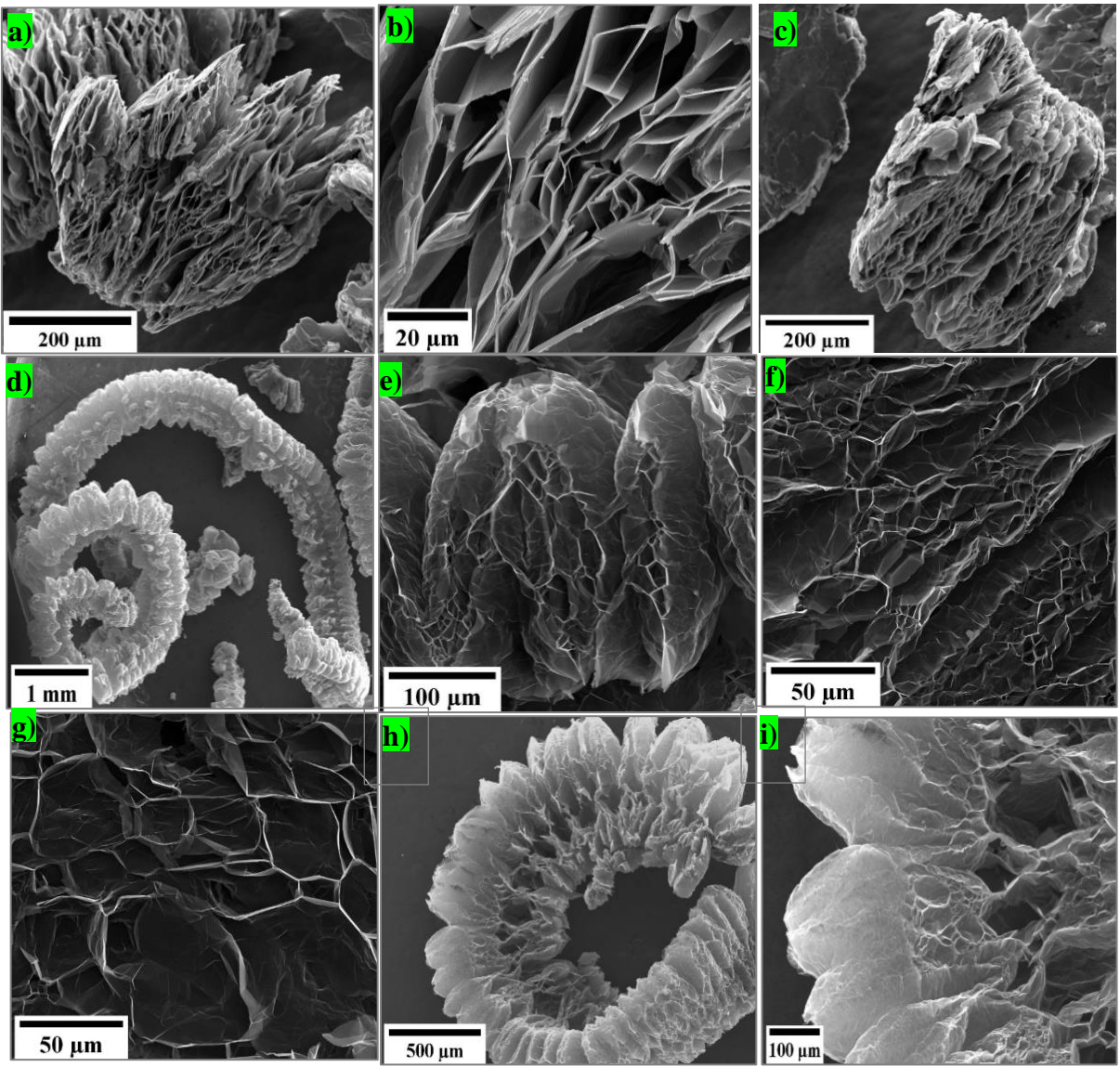

**Figure 19** FE-ESEM micrographs of a) GIC-H$_2$O$_2$ 1 (×150), b) GIC-H$_2$O$_2$ 1 (×800), c) GIC-H$_2$O$_2$ 1 (×120), d) EG-H$_2$O$_2$ 1 (×20), e) EG-H$_2$O$_2$ 1 (×250), f) EG-H$_2$O$_2$ 1 (×500), g) EG-H$_2$O$_2$ 1 (×500), h) EG-H$_2$O$_2$ 1 (×50), and i) EG-H$_2$O$_2$ 1 (×150).

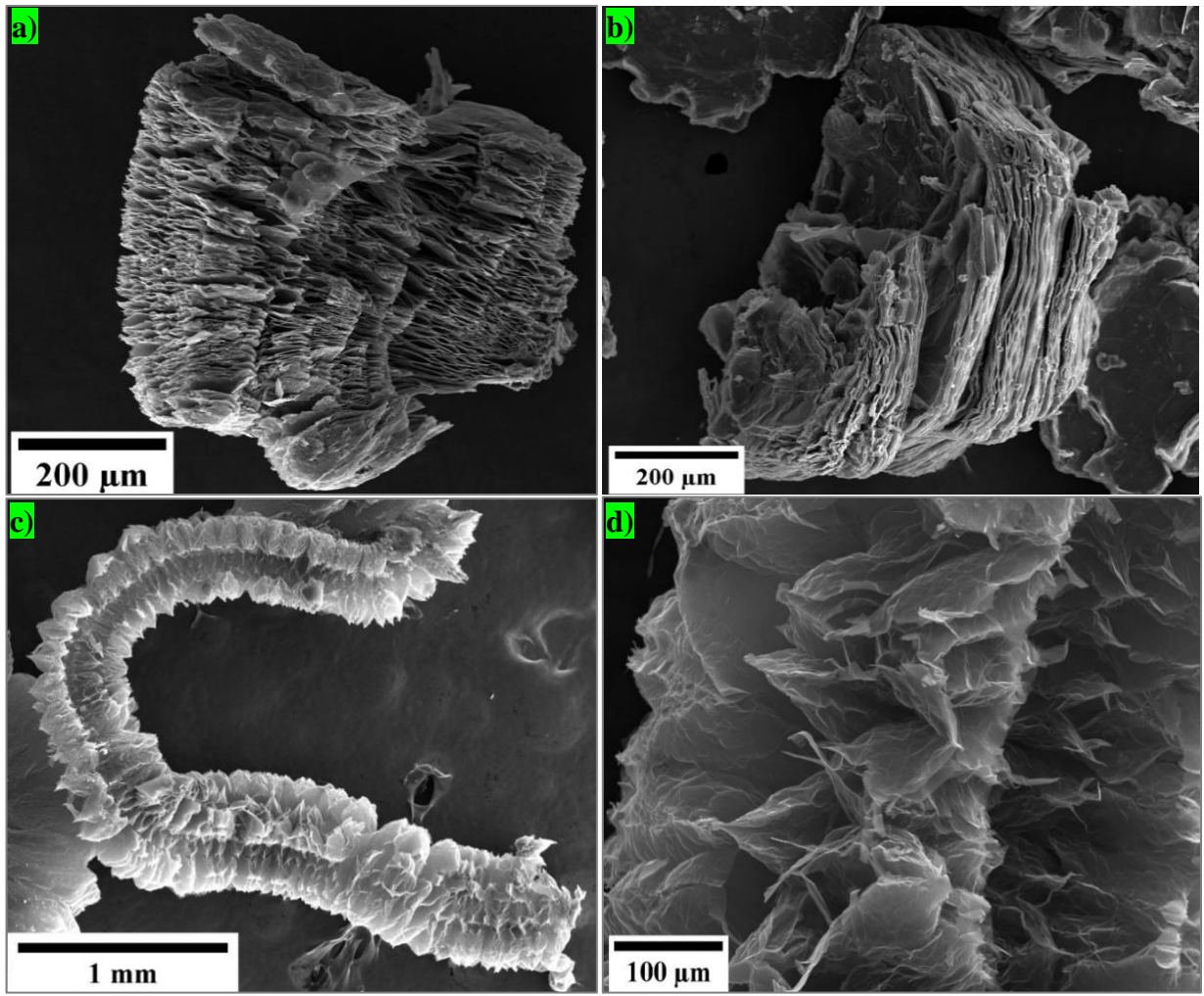

**Figure 20** FE-ESEM micrographs of a) GIC-$H_2O_2$ 2 (×120), b) GIC-$H_2O_2$ 3 (×120), c) EG-$H_2O_2$ 3 (× 35), d) EG-$H_2O_2$ 3 (×200).

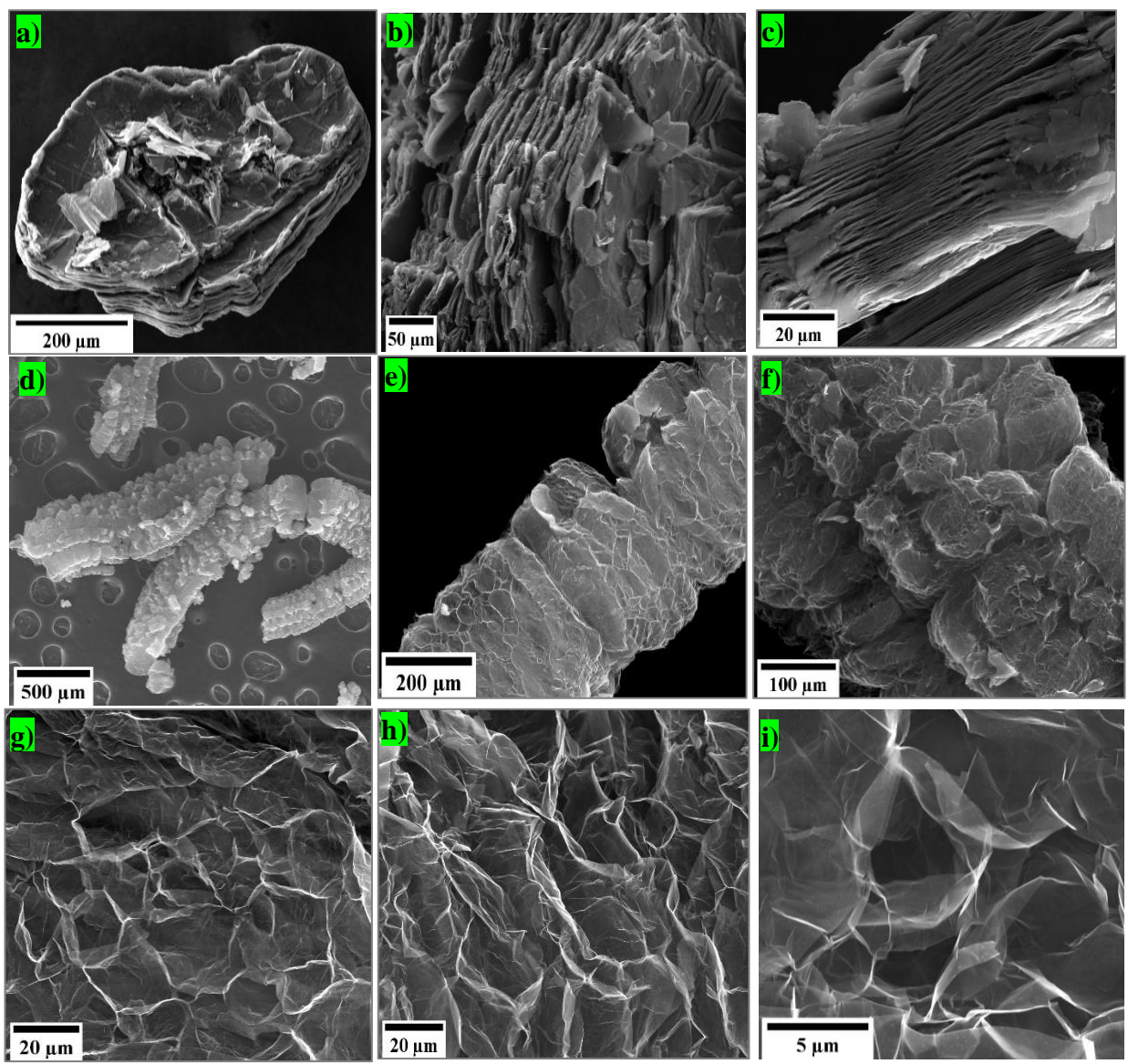

**Figure 21** FE-ESEM micrographs of a) GIC-NaClO$_3$ 1 (×150), b) GIC-NaClO$_3$ 1 (×250), c) GIC-NaClO$_3$ 1 (×1000), d) EG-NaClO$_3$ 1 (×35), e) EG-NaClO$_3$ 1 (×120), f) EG-NaClO$_3$ 1 (×500), g) EG-NaClO$_3$ 1 (×1000), h) EG-NaClO$_3$ 1 (×800), i) EG-NaClO$_3$ 1(×5000).

The EG-H₂O₂ 1 filler's edges are wide open (Figure 19 e), allowing the polymer to be absorbed properly into the EG-H₂O₂ 1 filler. This interconnected porous structure (seen in Figure 19 h & i), allows creation of an interpenetrating 3D polymer/graphene network in the composite. We have observed the effect on H₂O₂ intercalation route with a higher amount of H₂SO₄. As a result, we discovered that GIC-H₂O₂ 2 has a larger volume and a more ordered porous structure (Figure 20a), but EG-H₂O₂ 3 has a higher structural defect because of lateral breakage and delamination in GIC-H$_2$O$_2$ 3 (Figure 20b). Figure 20 c & d show the images of EG-H$_2$O$_2$ 3 after the thermal treatment.

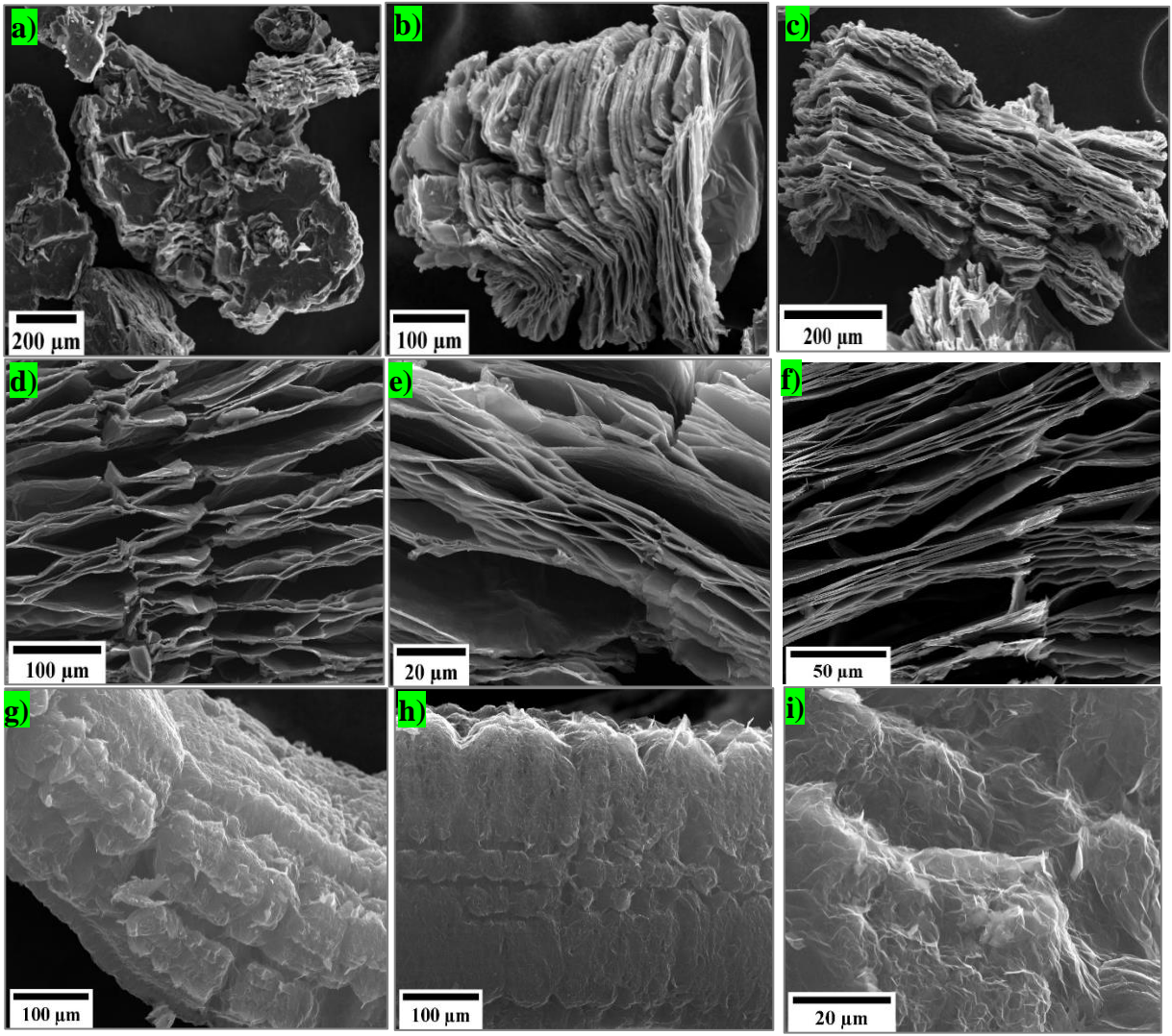

**Figure 22** FE-ESEM micrographs of a) GIC-NaClO₃ 2 (×80), b) GIC-NaClO₃ 2 (×200), c) GIC-NaClO₃ 2 (×150), d) GIC-NaClO₃ 2 (×250), e) GIC-NaClO₃ 2 (×500), f) GIC-NaClO₃ 2 (× 500), g) EG-NaClO₃ 2 (× 200), h) EG-NaClO₃ 2 (× 200), and i) EG-NaClO₃2 (×1200)

We have also observed the structural changes of GIC-NaClO$_3$ and EG-NaClO$_3$ while using NaClO$_3$ as an auxiliary intercalating agent to analyze the impact on the thermal properties of the composite. Figure 21a shows a fractured section at the center or basal plane area, as well as a crack across the GIC-NaClO$_3$ 1 particle. Figure 21b & c show highly stacked but delaminated area through the thickness, which can be explained as the result of vigorous intercalation.

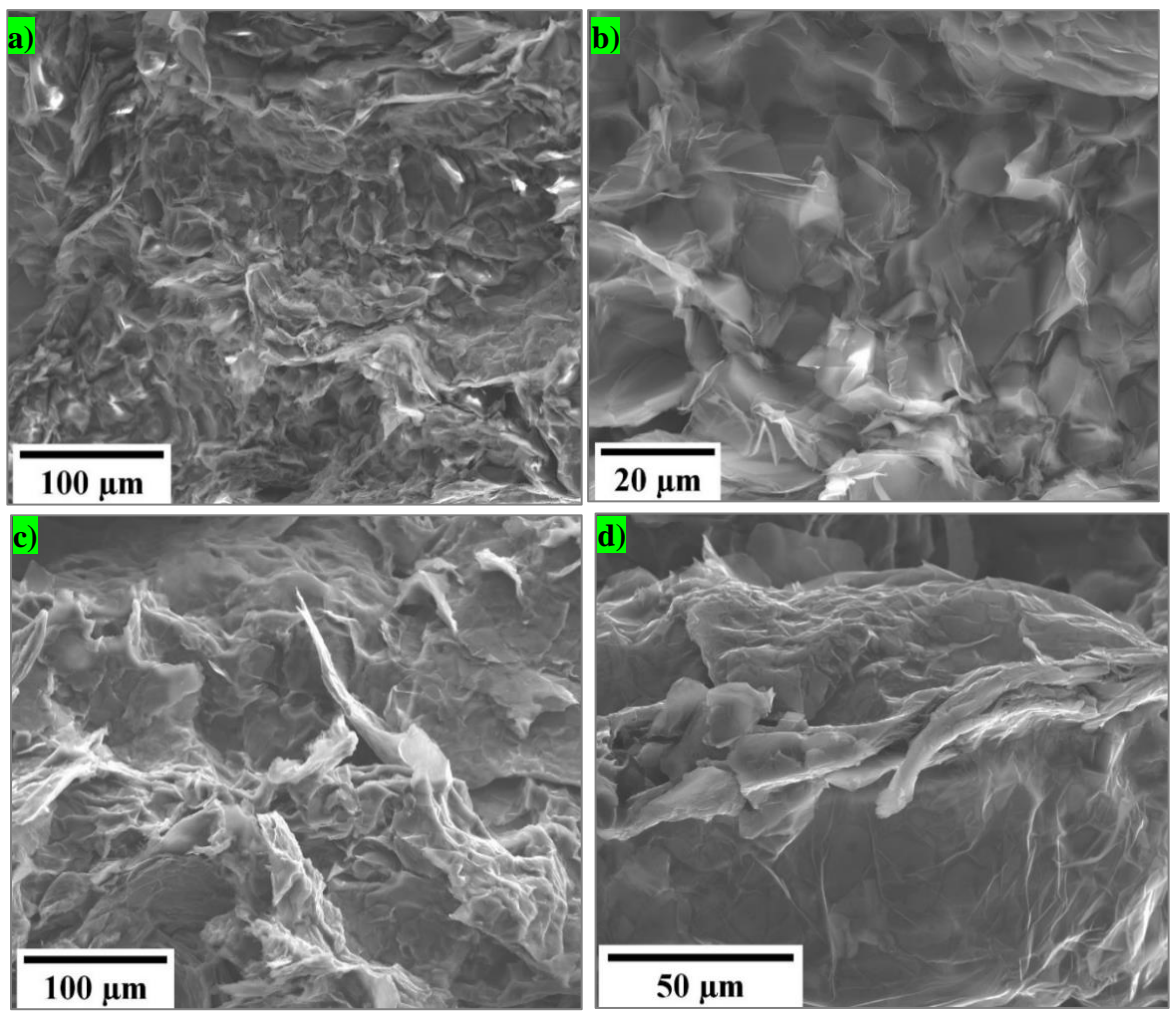

**Figure 23** FE-ESEM micrographs of EG-H$_2$O$_2$ 1/PEI composite for 10 wt% EG-H$_2$O$_2$ 1 filler a) (×250), b) (×650) magnification; and FE-ESEM micrographs of GIC-NaClO$_3$ 1/PEI composite for 10 wt% GIC-NaClO$_3$ 1 filler c) (×250), d) (×650) magnification.

The graphite layers expand when the intercalating chemicals transform into $CO_2$ and $SO_2$, and the length or surface of the GICs can be readily broken due to the loosely connected layers or fractured surface along the graphite layers (Figure 21 d). The expanded graphite image with the cleavage area because of the fractured nature is shown in Figure 21 e & f. Magnified view of comparatively smaller pores are visible in Figure 21 g-i.

We also observe the effect of longer intercalation time in the images of resulting GIC-$NaClO_3$ 2 and EG-$NaClO_3$ 2 in Figure 22 a-i. Figure 22 a, b & c represent the basal plane and cross-sectional area of a GIC -$NaClO_3$ 2 particle respectively. The structures seem to have a significant number of cracks and defects. Such structure leads to a reduction in the lateral size of the filler even after expansion.

To investigate the composite structure after composite fabrication, FE-ESEM images of EG-$H_2O_2$ 1/PEI composite (Figure 23 a & b) for 10 wt% EG-$H_2O_2$ 1 filler and EG-$NaClO_3$ 1/PEI composite (Figure 23 c & d) for 10 wt% GIC-$NaClO_3$ 1 filler were also observed. Because the pores of the EG-$H_2O_2$ 1 filler are filled with polymer, the surface appears smooth and homogenous, lowering the overall thermal resistance of polymer composites (Figure 23 a & b).

There are obvious differences between the two intercalated graphite's structure and thermally treated EG fillers. The important factor for EG-$H_2O_2$/PEI composite's higher thermal conductivity over EG-$NaClO_3$/PEI composite is insignificant structural flaws with the ordered texture of EG-$H_2O_2$.

## 5 CONCLUSIONS

In summary, two distinct intercalation methods have been employed to prepare GIC and EG filler using $H_2SO_4$ as a principal intercalator and two auxiliary intercalating agents, $H_2O_2$ and $NaClO_3$ as oxidizers. With optimized conditions, EG-$H_2O_2$ 1/PEI composites exhibit a $k$ value of 9.5 $Wm^{-1}K^{-1}$ when the EG-$H_2O_2$ 1 filler content is increased to 10 wt%, showing ~4030% enhancement relative to the pure PEI ($k$ ~0.23 $Wm^{-1}K^{-1}$). In comparison, the EG-$NaClO_3$ 1/PEI composite shows $k$ value of 3 $Wm^{-1}K^{-1}$ at 10 wt% EG-$NaClO_3$ 1 filler content, showing ~2190% enhancement relative to the PEI. Using the solution casting technique, the interconnected continuous network is established for the EG-$H_2O_2$ 1/PEI composites, whereas the continuous structure is not quite apparent in EG-$NaClO_3$ 1/PEI composites as compared by FE-ESEM analysis. The developed interconnected network of high-quality larger graphene nanosheet in EG-$H_2O_2$ relative to EG-$NaClO_3$ leads to superior composite thermal conductivity. Analysis of the morphology, structural integrity, and crystal structure through FE-ESEM, Raman and XRD has validated the superiority of the intercalation route involving $H_2O_2$. XPS analysis of the chemical composition of the products from the two intercalation routes reveals the preferential edge oxidation in EG-$H_2O_2$ 1, which leads to lower structural disorder relative to EG-$NaClO_3$ 1, caused by basal plane oxidation. Higher thermal diffusivity value of EG-$H_2O_2$ 1 filler than that of EG-$NaClO_3$ 1 filler further confirms the beneficial effect of the auxiliary intercalating agent, $H_2O_2$ used in the intercalation route I. With this superior thermally conductive composite, we have shed light on the potential use of EG filler in achieving an efficient thermal management system.

**Acknowledgment**

FT, SD, JG, and AN acknowledge support from National Science Foundation CAREER award under Award No. #1847129.

**Conflicts of Interest**

There are no conflicts to declare.